\pdfoutput=1

\documentclass[11pt,twoside,a4paper,cmspaper,final,collab]{cms-tdr}
\def\svnVersion{430537P}\def\cmsCernNoTag{CERN-EP-2017-085}\def\cmsCernDate{\today}\def\cmsMessage{Published in the Journal of High Energy Physics as \href{http://dx.doi.org/10.1007/JHEP10(2017)131}{\doi{10.1007/JHEP10(2017)131.}}}

\begin{document}\cmsNoteHeader{SMP-15-003}

\hyphenation{had-ron-i-za-tion}
\hyphenation{cal-or-i-me-ter}
\hyphenation{de-vices}
\RCS$HeadURL: svn+ssh://svn.cern.ch/reps/tdr2/papers/SMP-15-003/trunk/SMP-15-003.tex $
\RCS$Id: SMP-15-003.tex 419374 2017-08-03 08:19:22Z hinzmann $

\newlength\cmsFigWidth
\ifthenelse{\boolean{cms@external}}{\setlength\cmsFigWidth{0.85\columnwidth}}{\setlength\cmsFigWidth{0.4\textwidth}}
\ifthenelse{\boolean{cms@external}}{\providecommand{\cmsLeft}{top\xspace}}{\providecommand{\cmsLeft}{left\xspace}}
\ifthenelse{\boolean{cms@external}}{\providecommand{\cmsRight}{bottom\xspace}}{\providecommand{\cmsRight}{right\xspace}}
\newcommand{\x}{\ensuremath{\phantom{<}}}

\cmsNoteHeader{SMP-15-003}
\title{Measurements of jet charge with dijet events in pp collisions at $\sqrt{s}= 8\TeV$}

\date{\today}

\abstract{
Jet charge is an estimator of the electric charge of a quark, antiquark, or gluon initiating a jet. It is based on the momentum-weighted sum of the electric charges of the jet constituents. Measurements of three charge observables of the leading jet in transverse momentum \pt are performed with dijet events. The analysis is carried out with data collected by the CMS experiment at the CERN LHC in proton-proton collisions at $\sqrt{s}=8\TeV$ corresponding to an integrated luminosity of 19.7\fbinv. The results are presented as a function of the \pt of the leading jet and compared to predictions from leading- and next-to-leading-order event generators combined with parton showers. Measured jet charge distributions, unfolded for detector effects, are reported, which expand on previous measurements of the jet charge average and standard deviation in pp collisions.
}

\hypersetup{%
pdfauthor={CMS Collaboration},%
pdftitle={Measurements of jet charge with dijet events in pp collisions at sqrt(s) =8 TeV},%
pdfsubject={CMS},%
pdfkeywords={CMS, physics, jet charge, EW, jets}}
\maketitle

\section{Introduction}

High-momentum quarks and gluons (partons) produced at particle colliders form showers of hadrons, which can be clustered into jets to obtain information about the properties of the partons initiating the shower, and hence about the hard scattering causing the jets.
A jet is not a fundamental object, but a product of a jet clustering algorithm that depends on the choice of recombination scheme and parameters.
Jets can be initiated not only by single high-momentum colored partons, but also multiple partons from the decay of high-momentum top quarks, $\PW$, $\PZ$, and Higgs bosons, or new particles beyond the standard model.
At leading order (LO) in quantum chromodynamics (QCD), we can distinguish the type of partons that initiate jets and refer to them as quark jets, antiquark jets, or gluon jets.
To distinguish signal from background, or to characterize a new particle, it is often important to identify the object initiating a jet by means of the properties of the reconstructed particles that define the jet. In particular, the electric charge quantum number of the original parton from which a jet is initiated can be estimated from a momentum-weighted sum of the charges of the particles in the jet~\cite{fieldfeynman1978}.

The idea of estimating the charge of a parton from a jet-based observable has a long history.
The jet charge observable was suggested initially by Field and Feynman~\cite{fieldfeynman1978}. It was first measured in deep inelastic scattering at Fermilab~\cite{berge1980,berge1981}, CERN~\cite{allen1981,allen1982,albanese1984,barlag1982}, and Cornell~\cite{erickson1979} in an effort to understand models of quarks and hadrons. Among its applications were the identification of the charge of b quark jets~\cite{Abe1995, Braunschweig1990, Abreu1992, Decamp1991,Acton1992,Acton1994,Abreu1996,Abe:1999ds}, the $\PW$ boson charge discrimination~\cite{aleph1998,Abreu:2001rpa,Acciarri:1999kn,Abbiendi:2000ej}, as well as the determination of the charge of the top quark at the Tevatron~\cite{d02007,Aaltonen:2013sgl} and the CERN LHC~\cite{atlas2011}.

Recent theoretical calculations~\cite{waalewijn2012,krohn2013} motivate a more detailed estimation of jet charge and promote its use in new applications.
It has been shown that, despite the large experimental uncertainty in fragmentation functions, certain jet charge properties can be calculated independently of Monte Carlo (MC) fragmentation models.
Therefore, a jet charge measurement helps to further understand hadronization models and parton showers.
Studies of the performance and discrimination power of jet charge as well as comparisons of dijet, $\PW$+jets, and $\PQt\PAQt$ data with simulated pp collisions have been reported by the ATLAS~\cite{atlas2013} and CMS~\cite{cms2013} Collaborations.
A measurement of the average and standard deviation of the jet charge distribution as a function of the transverse momentum \pt of jets was recently published by the ATLAS~\cite{Aad:2015cua} Collaboration.

This paper presents a measurement of the jet charge distribution, unfolded for detector effects, with dijet events in pp collisions.
This result expands upon a previous work~\cite{Aad:2015cua} that reported the average and standard deviation of the jet charge distribution.
The measurement, performed in various ranges of \pt, is carried out for different definitions of jet charge to gain a better understanding of the underlying models that can be used to improve the predictions of MC event generators.

\section{The CMS detector}

The central feature of the CMS apparatus is a superconducting solenoid of 6\unit{m} internal diameter, providing a magnetic field of 3.8\unit{T}. A lead tungstate crystal electromagnetic calorimeter (ECAL) and a brass and scintillator hadron calorimeter (HCAL), each composed of a barrel and two endcap sections, reside within the solenoid volume. A preshower detector consisting of two planes of silicon sensors interleaved with lead is located in front of the ECAL at pseudorapidities $1.653 < |\eta| < 2.6$. An iron and quartz-fiber Cherenkov hadron calorimeter covers $3.0 < |\eta| < 5.0$. Muons are measured in gas-ionization detectors embedded in the steel flux-return yoke outside the solenoid.

Charged particle trajectories are measured with the silicon tracker within  $|\eta| < 2.5$. The tracker has 1440 silicon pixel and 15\,148 silicon strip detector modules. For nonisolated particles with $1 < \pt < 10\GeV$ and $\abs{\eta} < 1.4$, the track resolutions are typically 1.5\% in \pt and, respectively, 25--90 and 45--150\mum in the transverse and longitudinal impact parameters~\cite{Chatrchyan:2014fea}.

The ECAL and HCAL provide coverage up to $|\eta| = 3.0$. In the region $\abs{ \eta } < 1.74$, the HCAL cells have widths of 0.087 in $\eta$ and 0.087 radians in azimuth ($\phi$). In the $\eta$-$\phi$ plane, and for $\abs{\eta} < 1.48$, the HCAL cells map on to $5 {\times} 5$ ECAL crystals arrays to form calorimeter towers projecting radially outwards from the nominal interaction point. At larger values of $\abs{ \eta }$, the size of the towers increases and the matching ECAL arrays contain fewer crystals.
In the barrel section of the ECAL, an energy resolution of about 1\% is achieved for unconverted or late-converting photons in the tens of \GeVns energy range. The remaining barrel photons have a resolution of about 1.3\% up to $\abs{\eta} = 1$, rising to about 2.5\% at $\abs{\eta} = 1.4$. In the endcaps, the resolution of unconverted or late-converting photons is about 2.5\%, while the remaining endcap photons have a resolution between 3 and 4\%~\cite{CMS:EGM-14-001}. When combining information from the entire detector, the jet energy resolution amounts typically to 15\% at 10\GeV, 8\% at 100\GeV, and 4\% at 1\TeV, to be compared to about 40\%, 12\%, and 5\% obtained when the ECAL and HCAL alone are used. 

The first level (L1) of the CMS trigger system~\cite{Khachatryan:2016bia}, composed of special hardware processors, uses information from the calorimeters and muon detectors to select the most interesting events within a fixed time interval of 3.2\mus. The high-level trigger (HLT) processor farm further decreases the event rate from $\approx$100\unit{kHz} to less than 1\unit{kHz} before data storage. A more detailed description of the CMS detector, together with a definition of the coordinate system and kinematic variables, can be found in Ref.~\cite{Chatrchyan:2008zzk}.

\section{Data and simulated samples}

The data used in this analysis were recorded with the CMS detector in 2012 at the CERN LHC at a center-of-mass energy $\sqrt{s}=8\TeV$ and corresponds to an integrated luminosity of 19.7\fbinv. 
Events were collected with loose jet requirements, based on ECAL and HCAL information, at the L1 trigger.
An HLT requirement of at least one jet with transverse momentum $\pt > 320\GeV$ is imposed, based on information from all detector components, as described in detail in the following section.
This trigger is 99\% efficient for events with at least one jet reconstructed offline with $\pt > 400\GeV$.
\par The MC event generators \PYTHIA{6}.4.26~\cite{Sjostrand:2006za},  \PYTHIA{8}.205~\cite{bib:pythia8}, \POWHEG~v2~\cite{bib:Frixione:2007vw,bib:Alioli:2010xd,bib:Nason:2004rx}, and \HERWIG{++}~2.5.0~\cite{refId0} are used. \PYTHIA{6}, \PYTHIA{8}, and \HERWIG{++} are based on the LO matrix-elements combined with parton showers (PSs), while \POWHEG provides both LO and next-to-leading-order (NLO) matrix-element predictions~\cite{bib:POWHEG_Dijet}, which are combined with \PYTHIA{8} (\POWHEG+~\PYTHIA{8}) or \HERWIG{++} (\POWHEG+~\HERWIG{++}) PSs. These PS models, used to simulate higher-order processes, follow an ordering principle motivated by QCD. Successive radiation of gluons from a highly energetic parton is ordered using some specific variable, \eg, \pt or the angle of radiated partons with respect to the parent one. The two generators differ in the choice of jet-ordering technique, as well as in the treatment of beam remnants, multiple interactions, and the hadronization model. \PYTHIA{6} uses a \pt-ordered PS model. It provides a good description of parton emission when the emitted partons are close in $\eta$-$\phi$ space. The Z2$^\ast$ tune~\cite{Chatrchyan:2013gfi,Khachatryan:2015pea} is used for the underlying event description. It resembles the Z2 tune~\cite{Field:2012jv} except for the energy extrapolation parameter that is dependent on the choice of parton distribution function (PDF) set. Partons are hadronized using the Lund string model~\cite{Andersson:1983ia,Sjostrand:1984iu}.
\PYTHIA{8} is used with the CUETP8M1~\cite{Khachatryan:2015pea} tune, which employs the LO NNPDF2.3~\cite{nnpdf1,nnpdf2} parametrization of the PDFs.
\PYTHIA{8} is based on the same parton showering and hadronization models as \PYTHIA{6}.

The \HERWIG{++} program with the EE3C tune~\cite{Gieseke:2003rz} is based on a PS model that uses a coherent branching algorithm with angular ordering of the showers~\cite{Gieseke:2003rz}. The partons are hadronized using a cluster model~\cite{Webber:1983if}, and the multiple-parton interaction is simulated using an eikonal multiple parton scattering model~\cite{Gieseke:2003rz}.
The generated events from \PYTHIA{6} and \HERWIG{++} are passed through the CMS detector simulation based on \GEANTfour~\cite{Agostinelli:2002hh}.

\POWHEG is used to generate QCD multijet predictions at LO with the CTEQ6L1~\cite{cteq} PDF set, at NLO with the CT10~\cite{Lai:2010vv} NLO PDF set, and at NLO with the HERAPDF~1.5~\cite{Aaron:2009aa} NLO PDF set combined with the \PYTHIA{8} PSs. In addition, the \POWHEG calculation at NLO with CT10 NLO PDF set is combined with the \HERWIG{++} PSs.

\section{Event reconstruction and event selection}

Jets are reconstructed from particle-flow (PF) candidates~\cite{Sirunyan:2017ulk} using the anti-\kt clustering algorithm~\cite{Cacciari:2008gp,Cacciari:2011ma} with a distance parameter $R = 0.5$. The PF algorithm identifies electrons, muons, photons, charged hadrons and neutral hadrons through an optimized combination of information from all subdetectors. Jets are clustered from the PF objects and the total momenta of the jets are calculated by summing their four-momenta. To reduce the contamination from additional pp interactions (pileup), charged particles emanating from other pp collision vertices are removed before clustering. Because of the nonuniform and nonlinear response of the CMS calorimeters, the reconstructed jets require additional energy corrections that are based on high-\pt jet events generated with \PYTHIA{6}~\cite{Sjostrand:2006za}. Corrections using in situ measurements of dijet, $\PGg$+jet, and $\PZ$+jet events~\cite{Chatrchyan:2011ds} are applied to measured jets to account for discrepancies with the MC simulated jets.
\par Events are selected by requiring at least two jets that pass the following selection criteria: The jets with leading and subleading \pt must lie within $| \eta | < 1.5$ and have $\pt > 400\GeV$ and $\pt > 100\GeV$, respectively. Events with spurious jets from noise and noncollision backgrounds are rejected by applying a set of jet identification criteria~\cite{CMS-PAS-JME-10-003}. Additional selection criteria are also applied to reduce beam backgrounds and electronic noise. At least one reconstructed primary vertex within a 24\unit{cm} window along the beam axis is required.
In the presence of more than one vertex that passes these requirements, the primary interaction vertex is chosen to be the one with the highest total $\pt^2$, summed over all the associated tracks.
The missing transverse momentum in the event $\pt^{\text{miss}}$ is defined as the magnitude of the vector sum of the \pt of all PF candidates, and we require that $\pt^{\text{miss}}$/$\sum{\pt}<$ 0.3 where $\sum{\pt}$ is the scalar sum of all PF candidates
After the event selection the data sample contains mainly QCD multijet events, while backgrounds are negligible.

\begin{figure}[htb]
\begin{center}
\includegraphics[width=8.0cm]{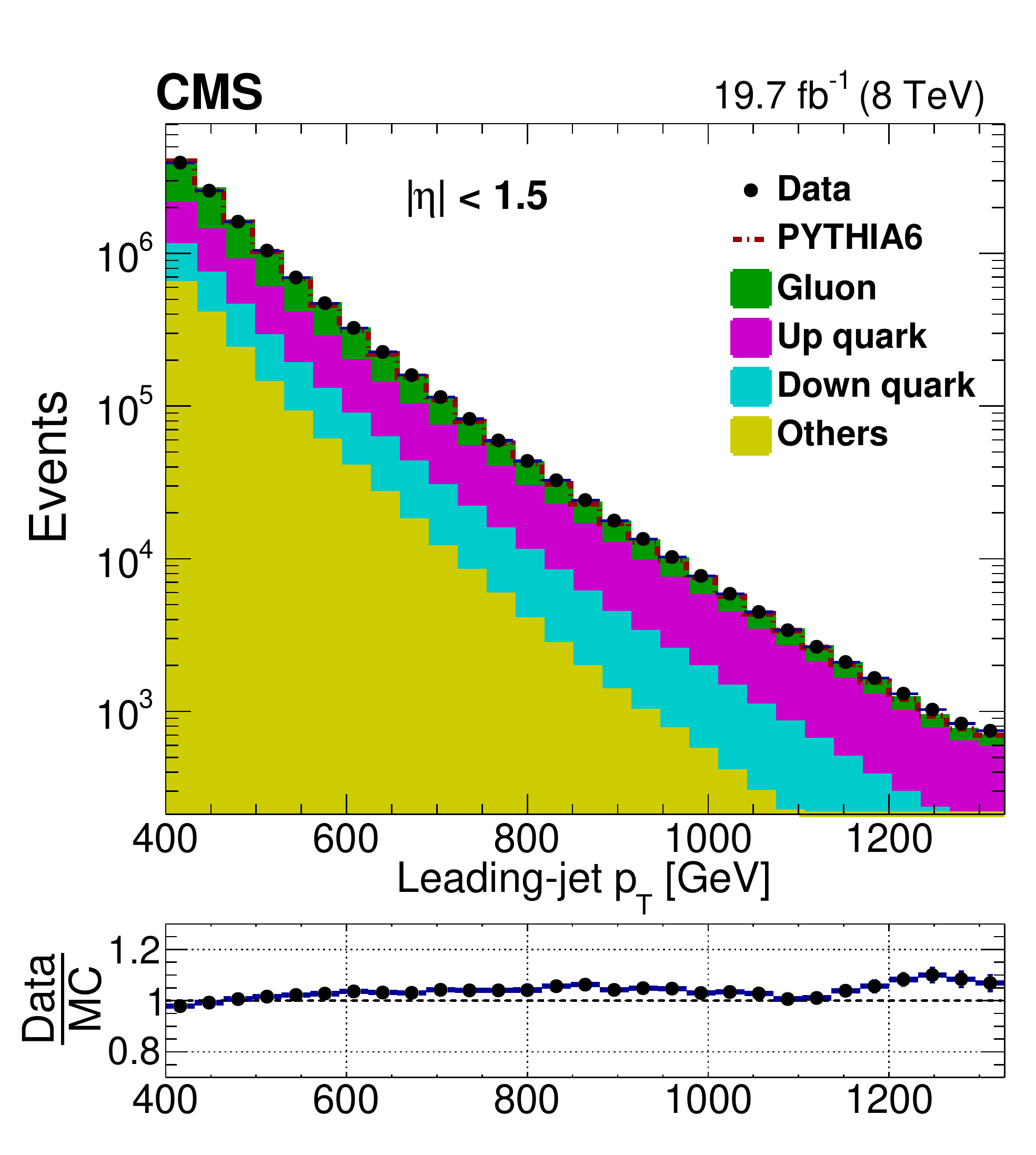}
\caption{\label{qgjets}Leading-jet \pt distribution in data (points) compared to \PYTHIA{6} simulation.
The \PYTHIA{6} prediction is normalized to match the total number of events observed in data.
Only statistical uncertainties are shown.
The filled histograms show the contributions from different types of initiating partons, identified by means of the matching algorithm described in the text.
The ``others'' category represents those jets that are initiated by up antiquark ($\cPaqu$), down antiquark ($\cPaqd$), charm, strange, and bottom (anti-)quarks (respectively, $\cPaqc$, $\cPqc$, $\cPaqs$, $\cPqs$, $\cPaqb$, $\cPqb$), and any unmatched jets.
The data points are shown in the center of each jet \pt bin.
}
\end{center}
\end{figure}

\par The agreement between data and MC simulations based on \PYTHIA{6} and \HERWIG{++} is verified at the reconstructed level using the kinematic properties of the leading jets: jet \pt, $\eta$, $\phi$, and dijet invariant mass, as well as jet properties, such as track multiplicity and jet charge.
Agreement at the 10\% level is found for each variable.
Figure~\ref{qgjets} provides a comparison of \PYTHIA{6} with the data as a function of the \pt of the leading jet.
For each \PYTHIA{6} event, the type of parton initiating the leading jet is identified with a geometrical matching procedure based on the distance $\Delta R$ in the $\eta$-$\phi$ plane between the generator-level hard partons and the reconstructed-level jet, where $\Delta R=\sqrt{\smash[b]{(\Delta \phi)^2+(\Delta \eta)^2}}$.
Before showering and radiation, the parton with the smallest $\Delta R$ with respect to the jet axis passing the matching criterion $\Delta R< \Delta R_{\text{max}}$, where $\Delta R_{\text{max}}$ = 0.3, is chosen as the parton initiating the jet.
Jets that cannot be matched to any generator-level hard parton with $\Delta R< \Delta R_{\text{max}}$ are categorized as unmatched. The matching efficiency is better than 96\% throughout the jet \pt range studied. The ``others'' category in Fig.~\ref{qgjets} represents those jets that are initiated by up antiquark ($\cPaqu$), down antiquark ($\cPaqd$), charm, strange, and bottom (anti-)quarks (respectively, $\cPaqc$, $\cPqc$, $\cPaqs$, $\cPqs$, $\cPaqb$, $\cPqb$), and any unmatched jets.

\section{Jet charge observables}
\label{sec-definitions}

Jet charge refers to the \pt-weighted sum of the electric charges of the particles in a jet.
Three definitions of jet charge are studied in this paper:
\begin{equation} \label{Omdef}
Q^\kappa = \frac{1}{(\pt^{\text{jet}})^\kappa} \sum_{i} Q_i (\pt^i)^\kappa ,
\end{equation}
\begin{equation}
Q_{L}^\kappa  =  \left.\sum_{i} Q_{i} \left(p^{i}_{\|}\right)^{\kappa} \middle/ \sum_{i} \left(p^{i}_{\|}\right)^{\kappa} \right. ,
\end{equation}
\begin{equation}
Q_{T}^\kappa  =  \left. \sum_{i} Q_{i} \left(p^{i}_{\perp}\right)^{\kappa} \middle/ \sum_{i} \left(p^{i}_{\perp}\right)^{\kappa} \right. .
\end{equation}
The first (``default'') definition follows Refs.~\cite{waalewijn2012,krohn2013}.
The sums above are over all color-neutral (electrically charged and neutral) particles $i$ in the jet that have $\pt > 1\GeV$.
The variable $\pt^{\text{jet}}$ is the transverse momentum of the jet, $Q_i$ is the charge of the particle, and $\pt^i$ is the magnitude of the transverse momentum of the particle relative to the beam axis.
In the $Q_{L}^\kappa$ (``longitudinal'') and $Q_{T}^\kappa$ (``transverse'') definitions,
the notations $p^{i}_{\|}  =  \vec{p}^{i}\cdot \vec{p}_{\text{jet}} / | \vec{p}_{\text{jet}}|$ and $p^{i}_{\perp} = | \vec{p}^{i}\times \vec{p}_{\text{jet}} | / | \vec{p}_{\text{jet}}|$ refer to the components of the transverse momentum of particle $i$ along and transverse to the jet axis, respectively.
The $\kappa$ parameter in the exponent of the particle momenta controls the relative weight given to low and high momentum particles contributing to the jet charge.
Values of $\kappa$ between 0.2 and 1.0 were used in previous experimental studies~\cite{berge1981,Decamp1991}.
Here three values of $\kappa$ are investigated: 0.3, 0.6, and 1.0.
The particle \pt cutoff of 1\GeV ensures that the dependence of the jet charge distributions on the number of pileup interactions in each event is negligible relative to the other sources of experimental uncertainty.

Compared to $Q^\kappa$, the quantity $Q_{L}^\kappa$ is more directly related to the fragmentation function $F(z)$ of a quark or a gluon,
which reflects the probability to find particle $i$ with momentum fraction $z = p^{i}_{\|}/|p_{\text{jet}}|$ in a quark jet or a gluon jet~\cite{fieldfeynman1978}.
We study all three variables $Q^\kappa$, $Q_{L}^\kappa$, and $Q_{T}^\kappa$ to elucidate the fragmentation of partons into hadrons.

At the generator level, the jet charge observables are computed in a similar way as above, using the generator-level stable particles (lifetime $\tau>10^{-12}\unit{s}$) with $\pt > 1\GeV$.

\begin{figure}[htb]
\begin{center}
\includegraphics[width=7.5cm]{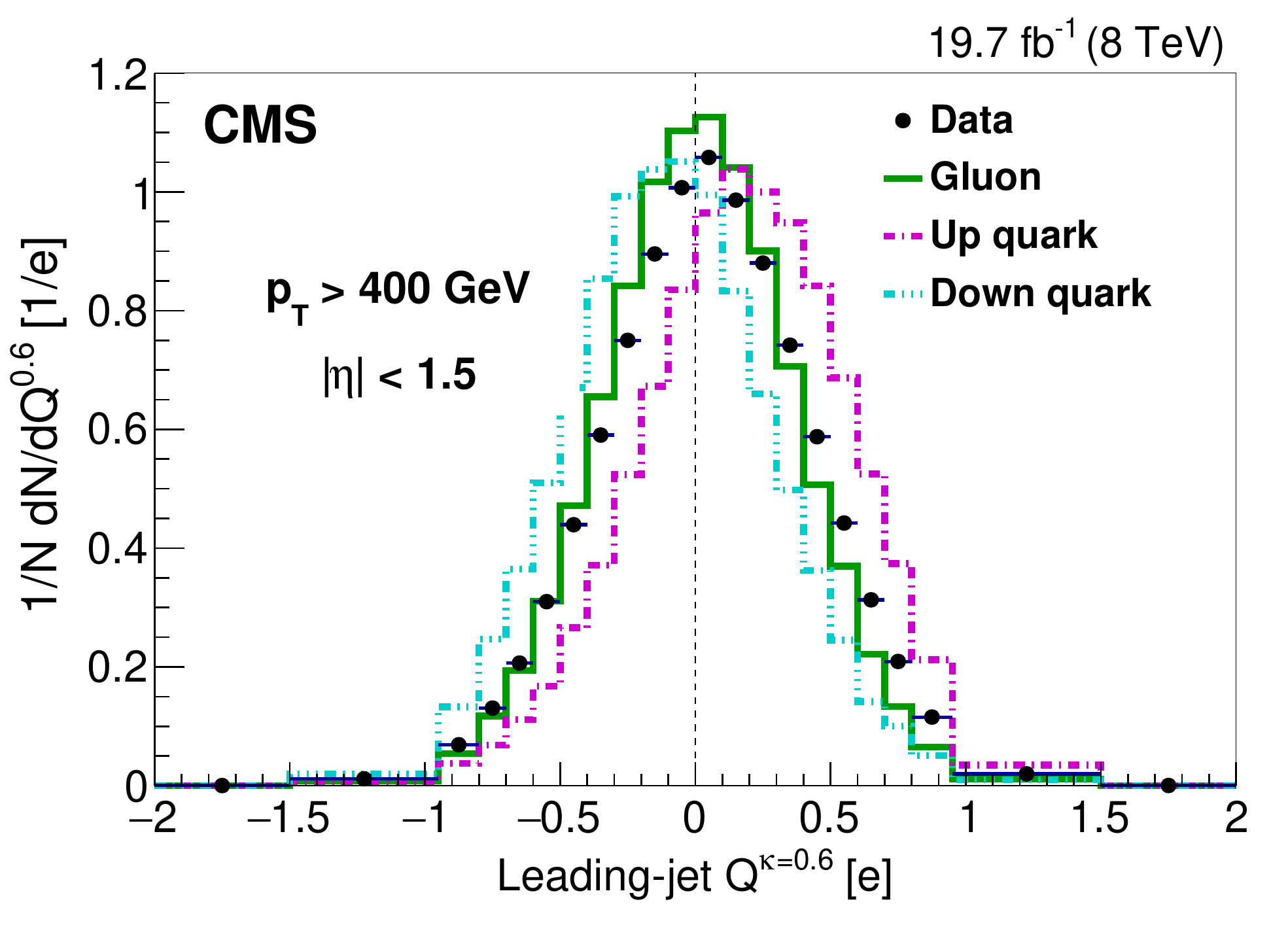}
\includegraphics[width=7.5cm]{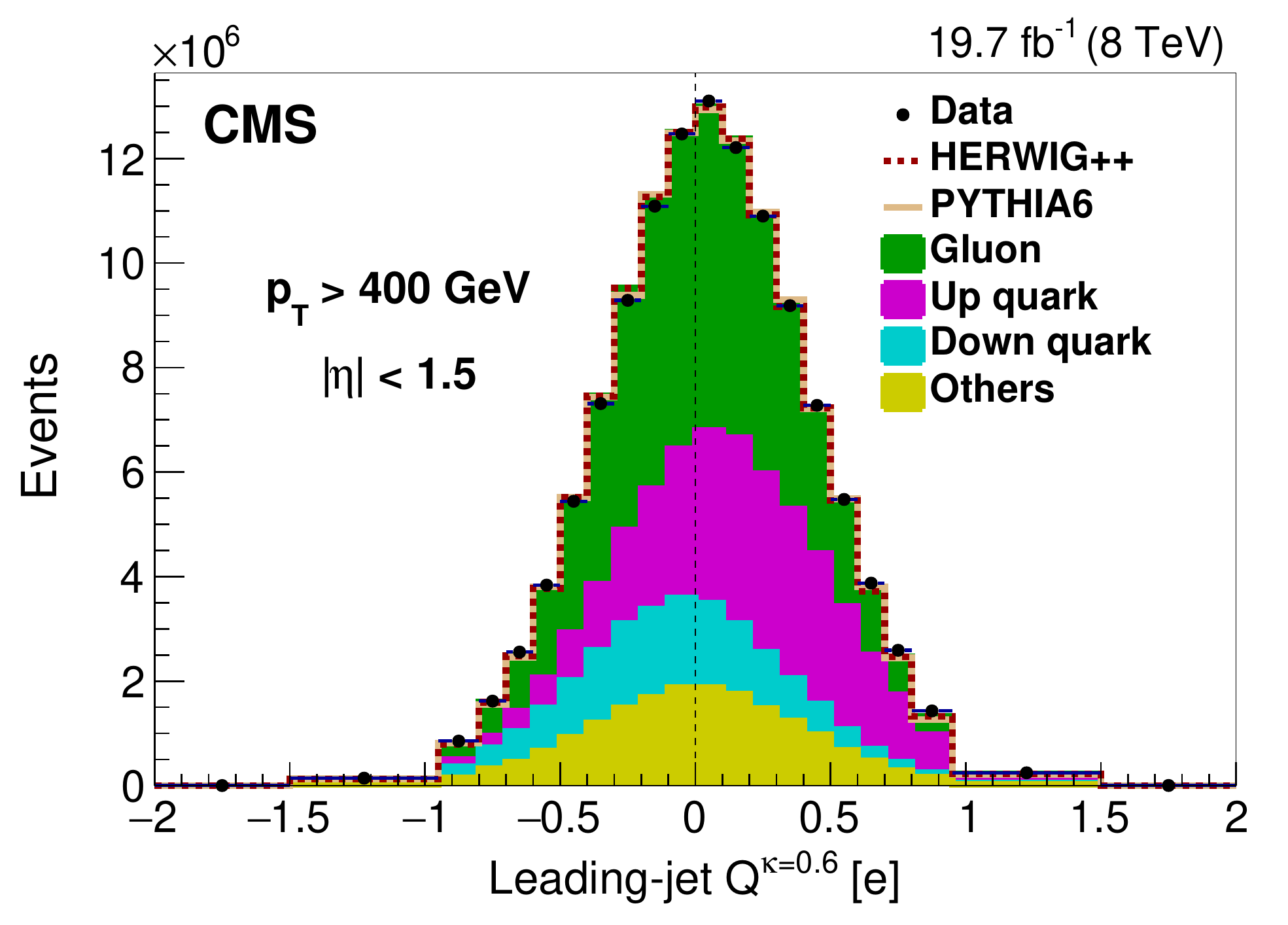} \\
\includegraphics[width=7.5cm]{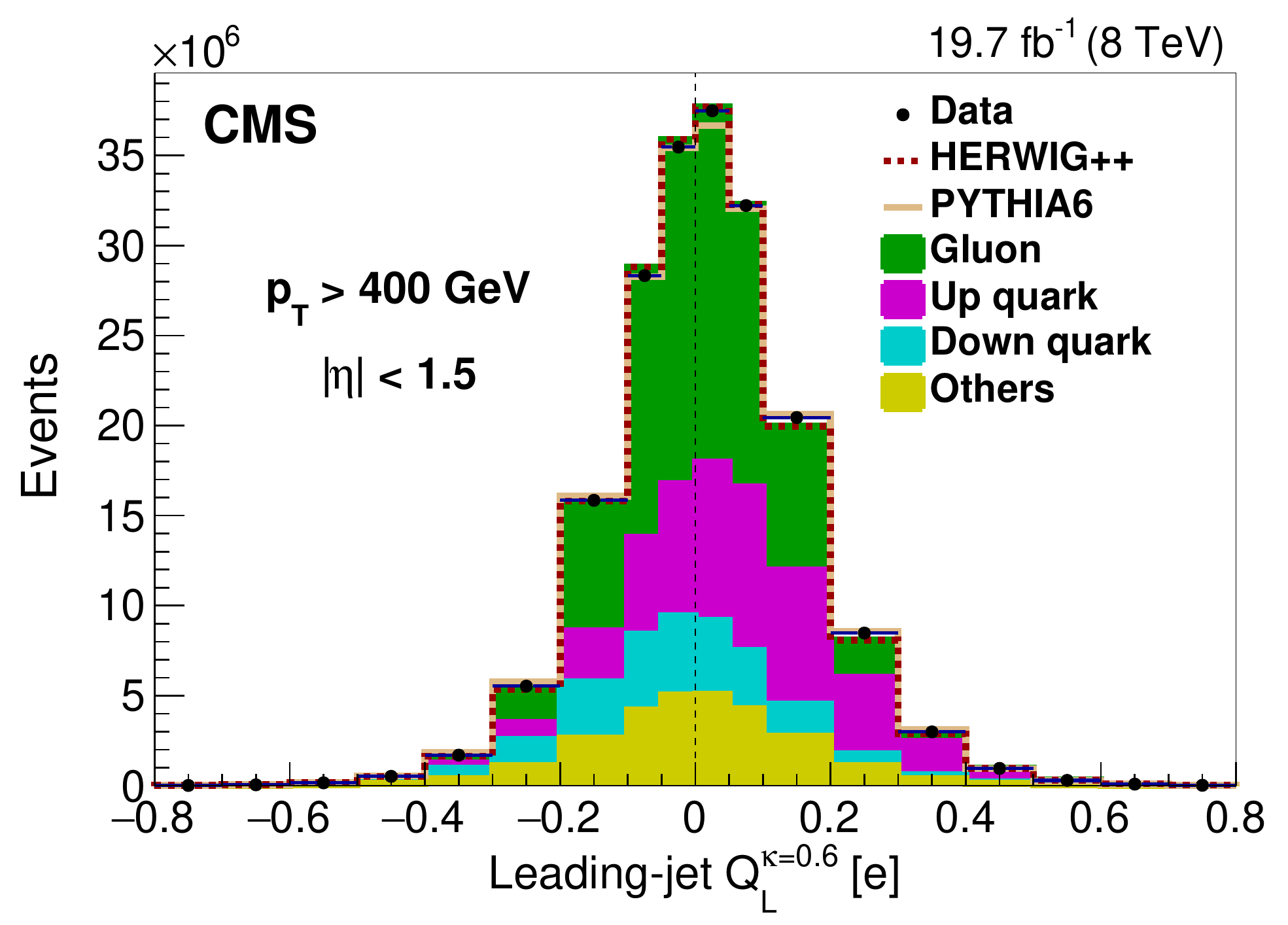}                                                                                                       
\includegraphics[width=7.5cm]{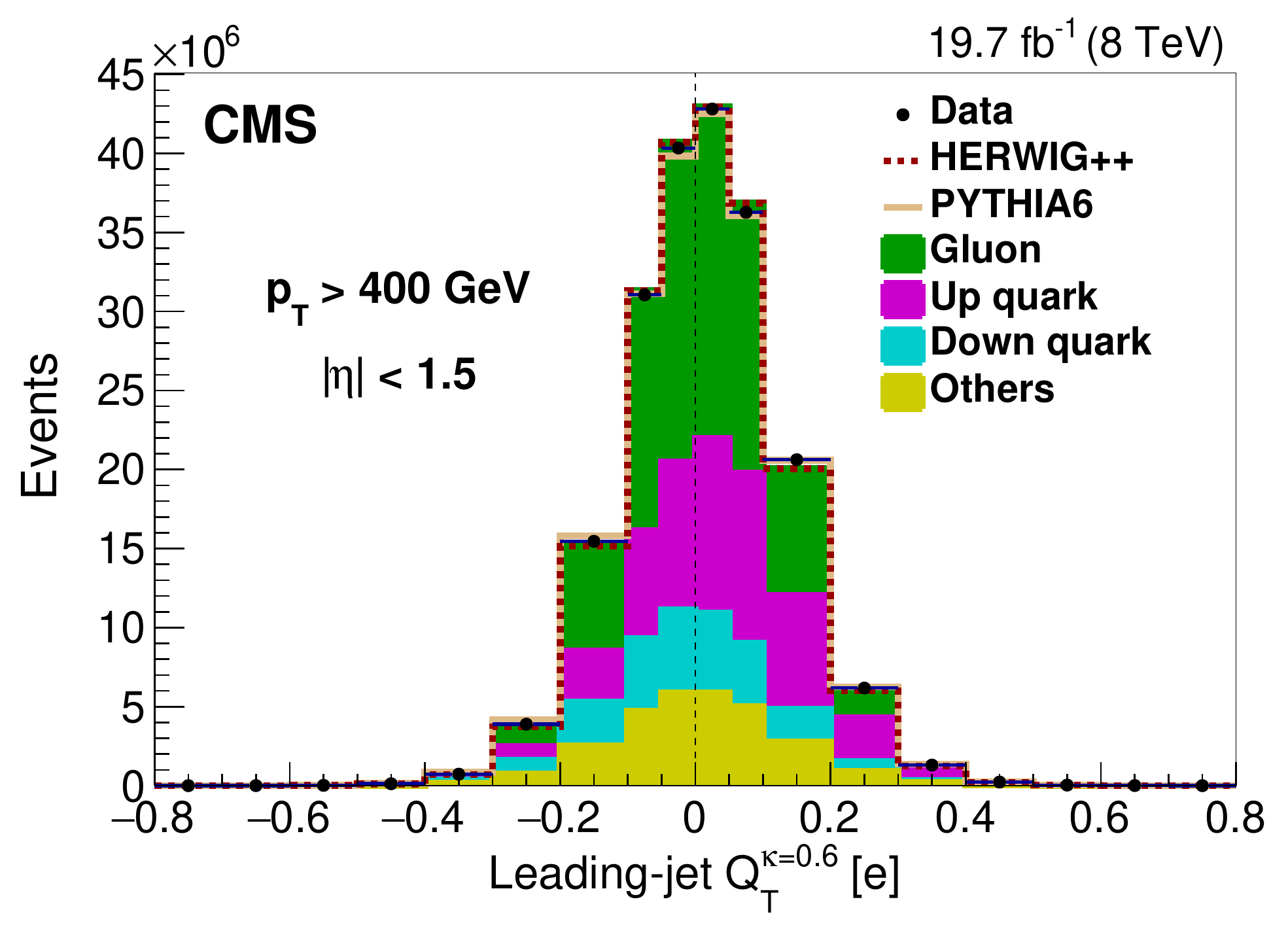}
\caption{\label{qgjets1}Distributions of jet charge for leading jets with $\kappa=0.6$ in data before unfolding (points) and MC simulations: $Q^\kappa$ (top row), $Q_{L}^\kappa$ (lower left), and $Q_{T}^\kappa$ (lower right).
The top left panel compares the data with the $\cPqu$, $\cPqd$, and $\cPg$ distributions from simulation based on \PYTHIA{6} where each distribution is normalized to unity.
The top right and lower panels compare the sum of the contributions in \PYTHIA{6} and \HERWIG{++} with data where each distribution is normalized to the observed number of data events. The parton assignment is determined from \PYTHIA{6}.
Only data statistical uncertainties are shown.
}
\end{center}
\end{figure}

Figure~\ref{qgjets1}~(upper left) compares data with the normalized charge distribution of the leading jet with $\kappa=0.6$, initiated by either an up quark ($\cPqu$), down quark ($\cPqd$), or a gluon (g) in \PYTHIA{6}.
The charge distribution for jets initiated by quarks with positive electric charge peaks at positive values, with a mean of $0.166e$,
as opposed to that for jets initiated by negatively charged quarks, with a mean of $-0.088e$ and gluons, with a mean of $0.013e$, where $e$ is the proton charge.
This suggests that the jet charge can be used to differentiate statistically jets from quarks of different electric charge, or to distinguish jets initiated by a gluon or a quark.
According to the simulated jet charge distribution shown in Fig.~\ref{qgjets1}~(upper left), $\approx$55\% of the down quark jets and $\approx$45\% of the gluon jets can be rejected at a selection efficiency of 70\% for up quark jets.

Figure~\ref{qgjets1}~(upper right and lower row) shows the jet charge data distribution compared with multijet predictions from \PYTHIA{6} and \HERWIG{++}, which are normalized to match the data.
Good agreement is observed between the data and the predictions from \PYTHIA{6} and \HERWIG{++}.
For \PYTHIA{6}, the prediction is broken down into contributions from different parton types.

\begin{figure}[htb]
\begin{center}
\includegraphics[width=8.0cm]{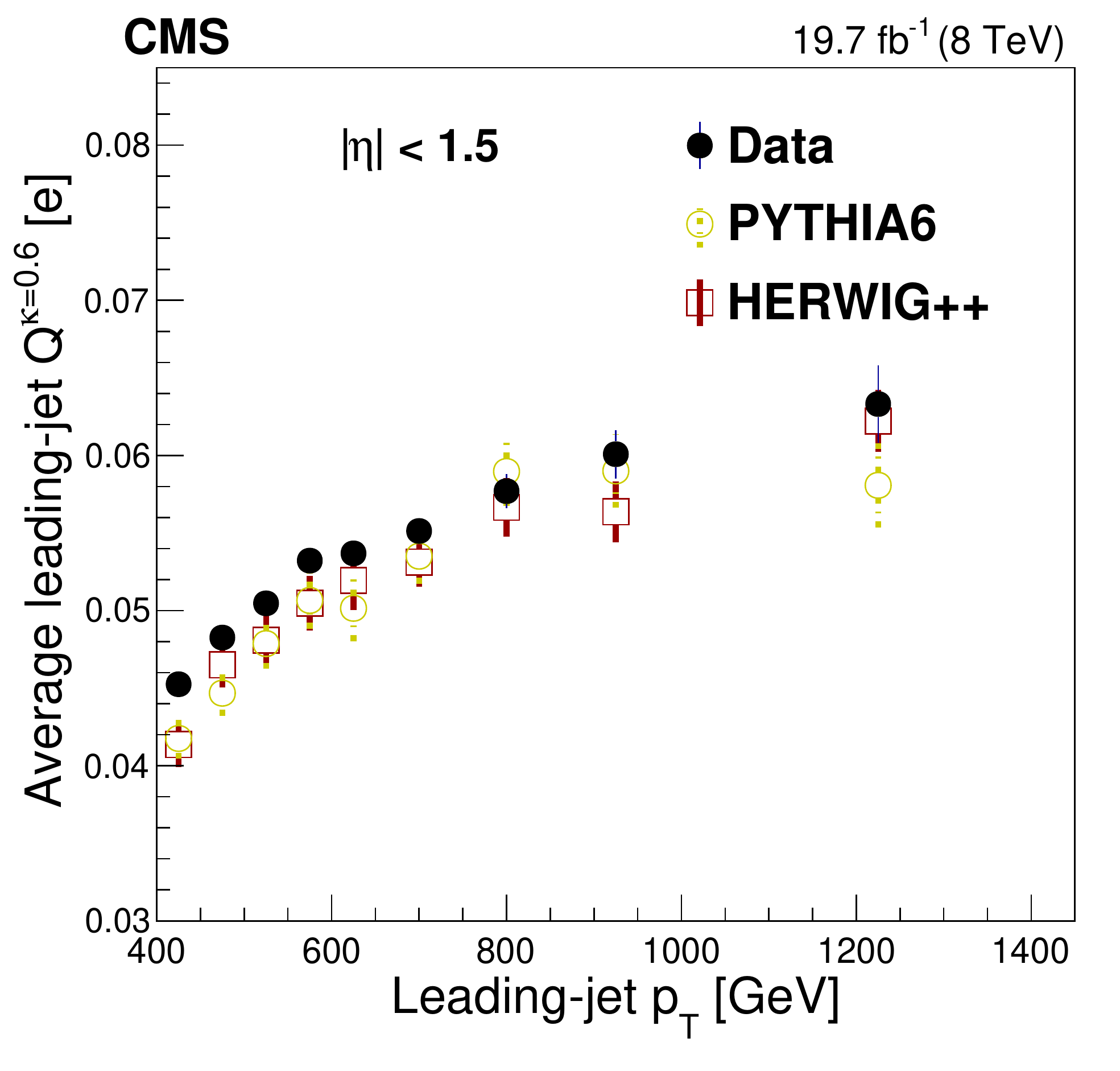}
\caption{\label{dijet4}
The data dependence of the average leading-jet charge $Q^\kappa$ with $\kappa=0.6$ on the \pt of the leading jet before unfolding and a comparison with simulations based on \PYTHIA{6} and \HERWIG{++}.
Only statistical uncertainties are shown.
The error bars for the simulation indicate the uncertainty from statistical fluctuations in the MC events.
The data points are shown in the center of each jet \pt bin. The bin boundaries are at 400, 450, 500, 550, 600, 650, 750, 850, 1000 and 1450\GeV.
}
\end{center}
\end{figure}

As shown in Fig.~\ref{qgjets}, the jet parton type composition of the selected dijet sample depends on the leading-jet \pt. Gluon jets dominate the lower part of the \pt spectrum, while up quarks become progressively more relevant at high \pt. As a consequence, the average jet charge with $\kappa=0.6$ increases as a function of the leading-jet \pt, as can be observed in Fig.~\ref{dijet4}. \PYTHIA{6} and \HERWIG{++} simulations reproduce this trend. It is therefore interesting to divide the dijet sample into different ranges of leading-jet \pt and measure the jet charge distribution separately in each subsample, thereby gaining information on the sensitivity of jet charge definitions to mixtures of parton types and the quality of the description offered by different generators.

\section{Unfolding of detector effects}

\begin{figure}[htb]
\begin{center}
\includegraphics[width=7.5cm]{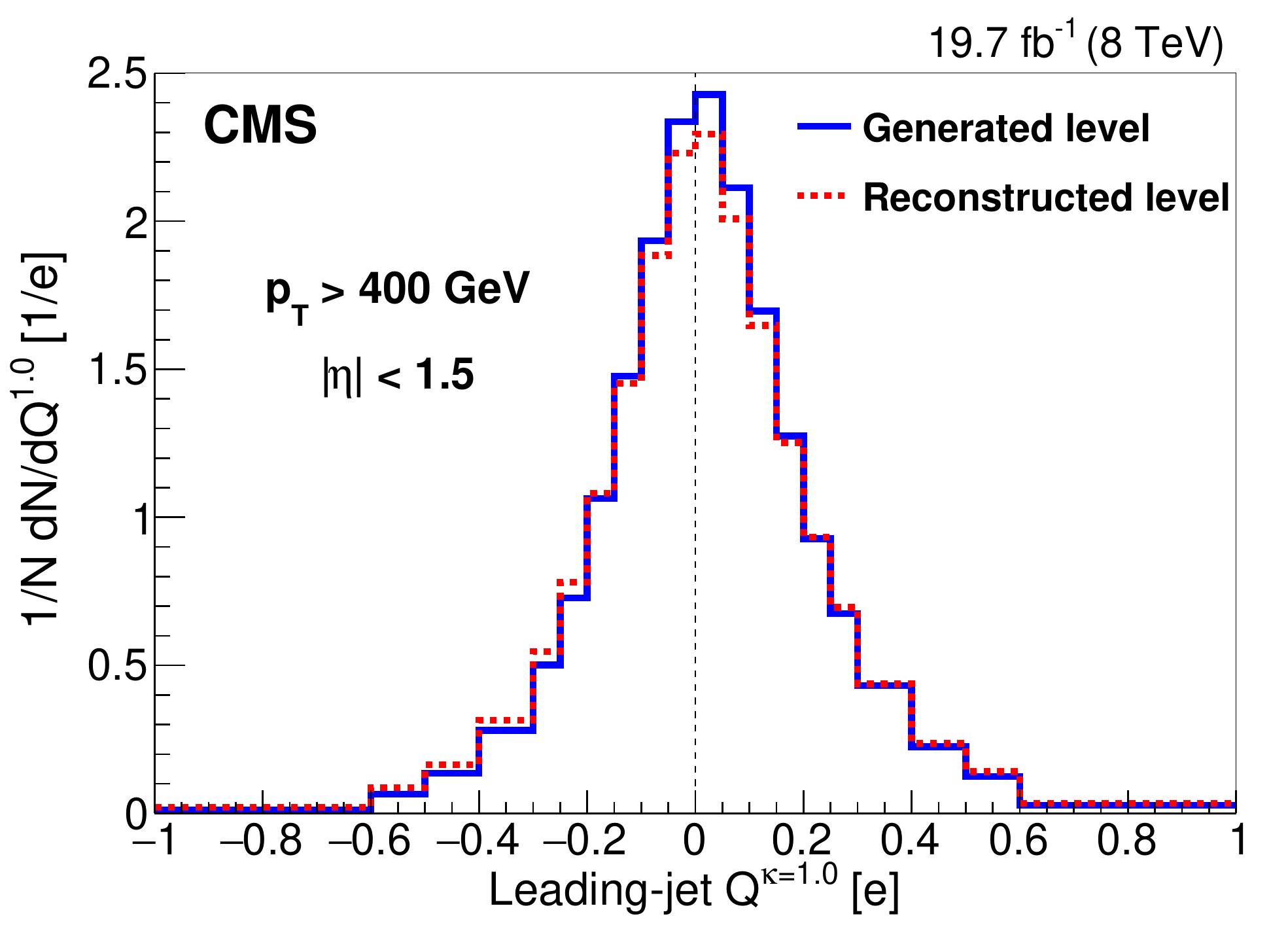}
\includegraphics[width=7.5cm]{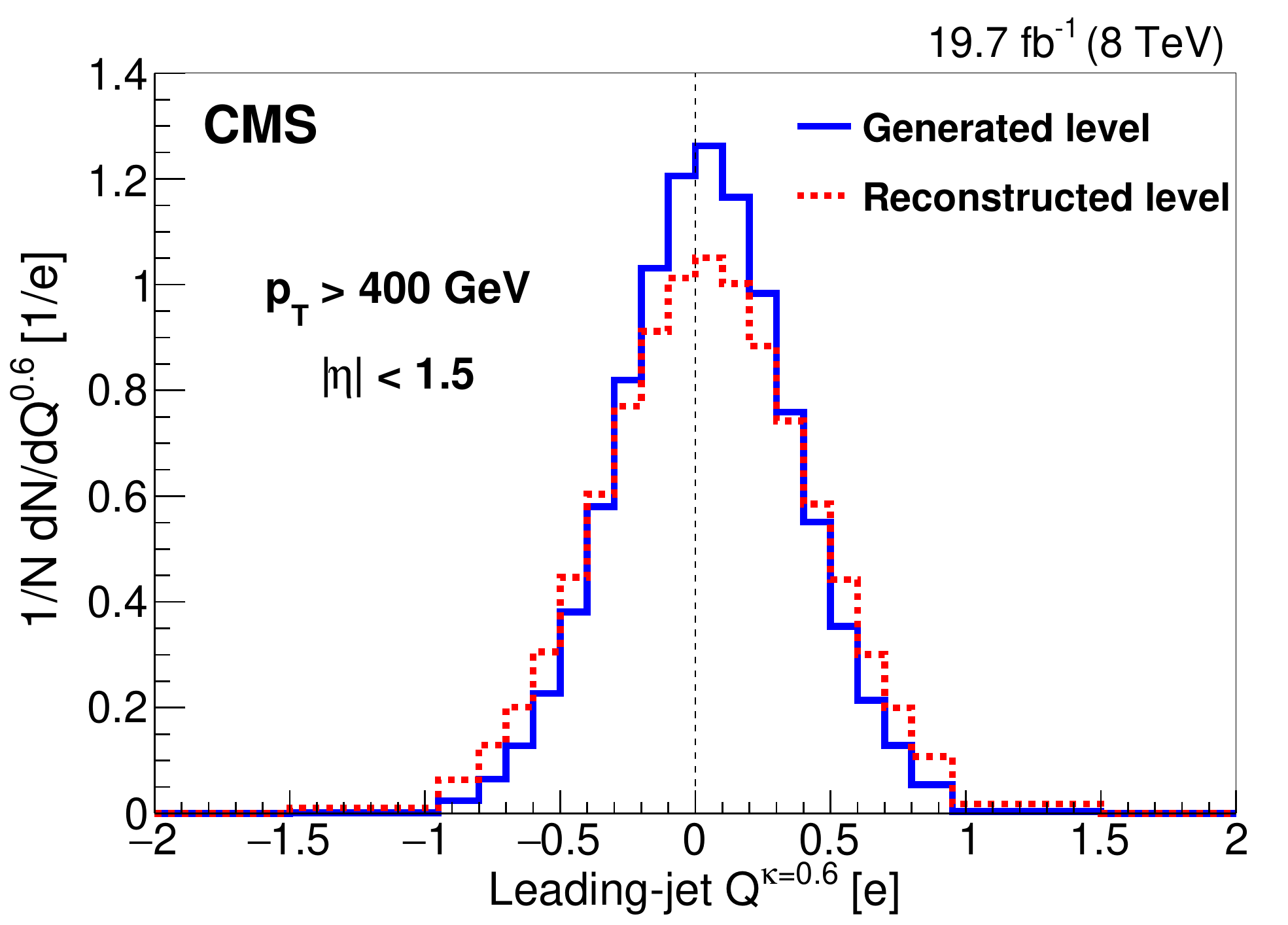}	\\									    
\includegraphics[width=7.5cm]{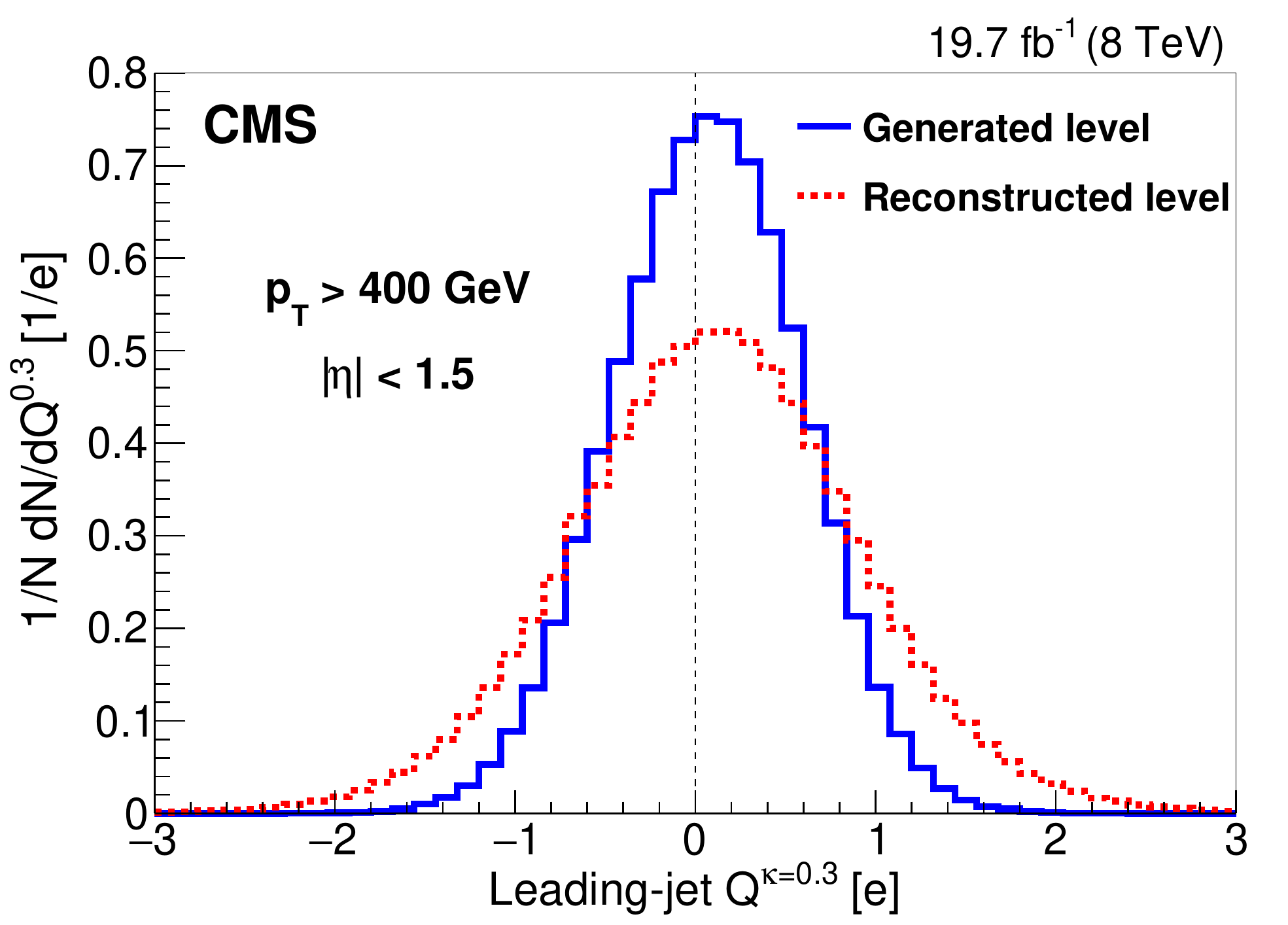}
\caption{\label{Unfoldeff}Distributions of leading-jet charge $Q^\kappa$ at the reconstructed level and generated levels in \PYTHIA{6}, for (upper left) $\kappa =$ 1.0, (upper right) 0.6, and (bottom) 0.3.}
\end{center}
\end{figure}

To compare with other measurements or theoretical predictions, the measured jet charge distributions must be unfolded from the resolution at the detector level to the final-state particle level. The jet charges in the MC simulation at the detector level are not identical to those constructed using the generator-level information, defined through some given theoretical input, because of detector resolution and acceptance effects. In particular, Fig.~\ref{Unfoldeff} shows that the difference between jet charge distributions at the generator level and the reconstructed level in \PYTHIA{6} increases with decreasing $\kappa$ values, because the definition of jet charge for small values of $\kappa$ gives more weight to low-\pt particles, which have a track reconstruction efficiency of about 90\%.

The unfolding is based on the D'Agostini iteration method with early stopping~\cite{DAgostini1995487,Richardson,Lucy}, where the unfolding utilizes a response matrix that maps the true onto the measured distribution. The response matrix is taken from the \PYTHIA{6} simulation and is used to unfold the data.  
The D'Agostini iteration method follows an iterative response-matrix inversion, in which the regularization is achieved by stopping the iteration just before the appearance of large fluctuations in the inverse matrix~\cite{DAgostini1995487}.
Another frequently used regularized unfolding algorithm, known as the singular value decomposition (SVD) method~\cite{Hocker:1995kb}, is utilized to cross-check the results. These two approaches agree roughly within about 0.7\%,
and both are implemented in the {\sc RooUnfold} software package~\cite{Adye:2011gm}.

\section{Systematic uncertainties}  

The experimental uncertainties that affect the measured results are summarized in this section.     
The uncertainties in jet energy scale and jet energy resolution are estimated by considering the corresponding effects in the computation of jet charge and then propagating the changes through the analysis.
The uncertainty in the jet energy scale is estimated to be 1--2.5\%~\cite{Chatrchyan:2011ds}, depending on the jet \pt and $\eta$. To map this uncertainty onto the jet charge variable, the reconstructed jet transverse momenta are systematically shifted by their respective uncertainty and the new values for the jet charge variables are calculated and compared.
The uncertainty in the momentum scale of the charged particles in a jet is negligible compared to the uncertainty in the jet energy scale and thus not varied.
The jet energy resolution is measured by comparing the asymmetry in the momenta of the two jets in dijet events~\cite{Chatrchyan:2011ds}.
The simulated jet energy resolution is smeared to match the measured resolutions and is changed by its uncertainty.

The jet charge is measured from the particles reconstructed from the charged tracks and calorimeter energy by the PF algorithm. For each track, the corresponding reconstruction efficiency varies with track \pt and $\eta$. The track reconstruction efficiency for charged pions is estimated in Ref.~\cite{Chatrchyan:2014fea} and is used as the weight factor for the PF objects. For each track, the corresponding track reconstruction efficiency is estimated, as a function of $\eta$ and \pt, from a simulated MC dijet event sample.
The resulting efficiency is varied by one standard deviation around its original value, and the jet charge variable is recalculated for each variation in the track weight factor.
The track \pt resolution depends on the track \pt and $\eta$. For example, the relative \pt resolution varies from 0.011 to 0.015 for a track \pt of about 1\GeV as $|\eta|$ changes from 0.5 to 1.0~\cite{Chatrchyan:2014fea}. For each track, the corresponding \pt resolution is estimated as a function of $\eta$ and \pt from a simulated MC dijet event sample.
The resulting resolution is then varied by one standard deviation of its original value, and the jet charge is computed for each change in track-\pt smearing. The jet energy scale and jet energy resolution have negligible correlations with track \pt resolution and track reconstruction efficiency.

To study the systematic effect arising from the choice of the \PYTHIA{6} generator to produce the response matrix used in the unfolding procedure, a response matrix is formed using \HERWIG{++}, and both of these matrices are used to unfold the data.
The corresponding difference is taken as the uncertainty in the modeling of the response matrix.
Another systematic effect taken into account in the unfolding procedure is the statistical uncertainty in the MC simulation of the matrix elements in the response matrix. They are propagated using the {\sc RooUnfold} software package.

The systematic uncertainty related to the modeling of pileup is estimated by comparing the jet charge distributions with varied pileup reweighting applied to the simulated samples within the uncertainty of the pileup distribution.
Table~\ref{tabsys} summarizes the sizes of the various systematic effects.
The impact of systematic effects on the jet charge distribution can be summarized by the quantity
\begin{equation} \label{errordef}
 \left. \sum_{i}{\frac{N_i^2}{\sigma_{N_i}^2} \frac{|N_i^{\text{upward}}-N_i^{\text{downward}}|}{N_i}}  \middle/ {\sum_{i}{\frac{N_i^2}{\sigma_{N_i}^2}}} \right. ,
\end{equation}
where the sums are over the bins $i=1,...,n_{\text{bins}}$ in the jet charge distribution, $N_i^{\text{up}}$ and $N_i^{\text{down}}$ are the respective one-standard-deviation upward and downward systematic changes in the nominal jet charge distribution $N_i$, and $\sigma_{N_i}$ is the statistical uncertainty in bin $i$ of the jet charge distribution.
The dominant uncertainties arise from the track \pt resolution and the modeling of the response matrix.
The remaining systematic uncertainties have small effects (less than a percent) and include the jet energy scale and jet energy resolution.
The jet charge computations for all three $\kappa$ values show comparable systematic uncertainties.

\begin{table}[htbp!]
\renewcommand{\arraystretch}{1.1}
\topcaption{\label{tabsys}Systematic uncertainties in terms of their corresponding inverse-variance-weighted mean in the fractional deviation as defined in Eq.~(\ref{errordef}) in percent (\%).}
\begin{center}
\resizebox{\textwidth}{!}{
\begin{tabular}{c|c|c|c|c|c|c|c|c|c} 
\hline
\multirow{2}{*}{Sources of uncertainty} & \multicolumn{3}{c|}{$\kappa =$ 1.0} & \multicolumn{3}{c|}{$\kappa =$ 0.6} & \multicolumn{3}{c}{$\kappa =$ 0.3} \\ 
& $Q^\kappa$ & $Q_{L}^\kappa$ & $Q_{T}^\kappa$ & $Q^\kappa$ & $Q_{L}^\kappa$ & $Q_{T}^\kappa$ & $Q^\kappa$ & $Q_{L}^\kappa$ & $Q_{T}^\kappa$ \\ \hline
Jet energy scale & 0.7 &$<$0.1\x &$<$0.1\x & 0.4 &$<$0.1\x &$<$0.1\x & 0.3&$<$0.1\x &$<$0.1\x \\ 
Jet energy resolution & 0.1 &$<$0.1\x&$<$0.1\x & 0.1 &$<$0.1\x&$<$0.1\x &$<$0.1\x &$<$0.1\x&$<$0.1\x \\ 
Track reconstruction & 0.4 & 0.4 & 0.5 & 0.5 & 0.4 & 0.5 & 0.5 & 0.4 & 0.4 \\ 
Track \pt resolution & 1.4 & 1.0 & 0.8 & 1.0 & 0.6 & 0.7 & 1.5 & 0.4 & 0.4 \\ 
Pileup & $<$0.1\x & $<$0.1\x & $<$0.1\x & $<$0.1\x & $<$0.1\x & $<$0.1\x & $<$0.1\x & $<$0.1\x & $<$0.1\x \\ 
Response matrix modeling & 1.6 & 1.6 & 1.8 & 1.0 & 0.8 & 1.3 & 1.5 & 1.3 & 1.3 \\ 
Response matrix statistics & 0.9 & 0.9 & 0.6 & 0.6 & 0.6 & 0.5 & 0.6 & 0.5 & 0.4 \\ \hline
\end{tabular}
}
\end{center} 
\label{tab:dataset}
\end{table}

\section{Results}

\begin{figure}[htb]
\begin{center}
\includegraphics[width=6.7cm]{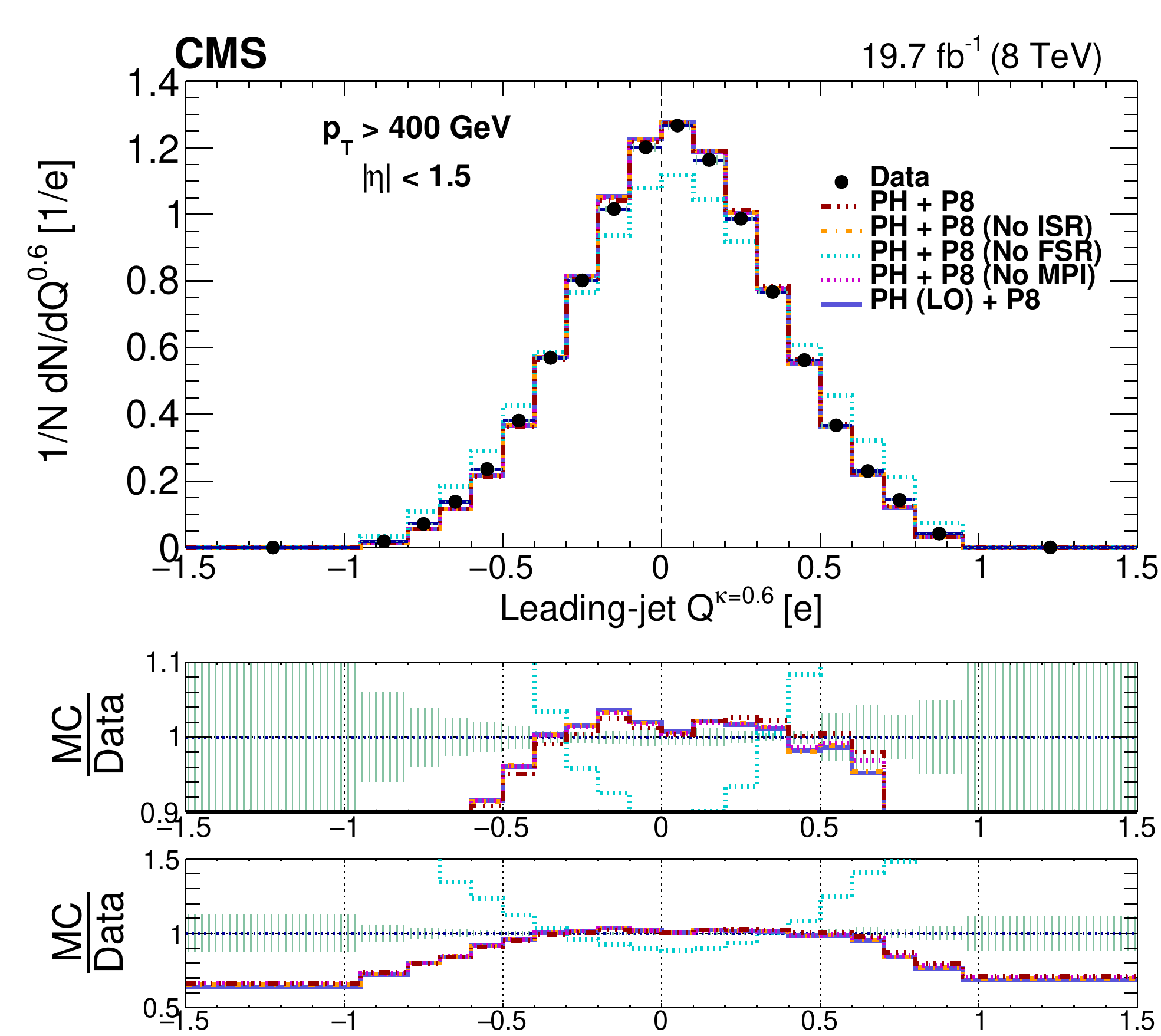} \\
\includegraphics[width=6.7cm]{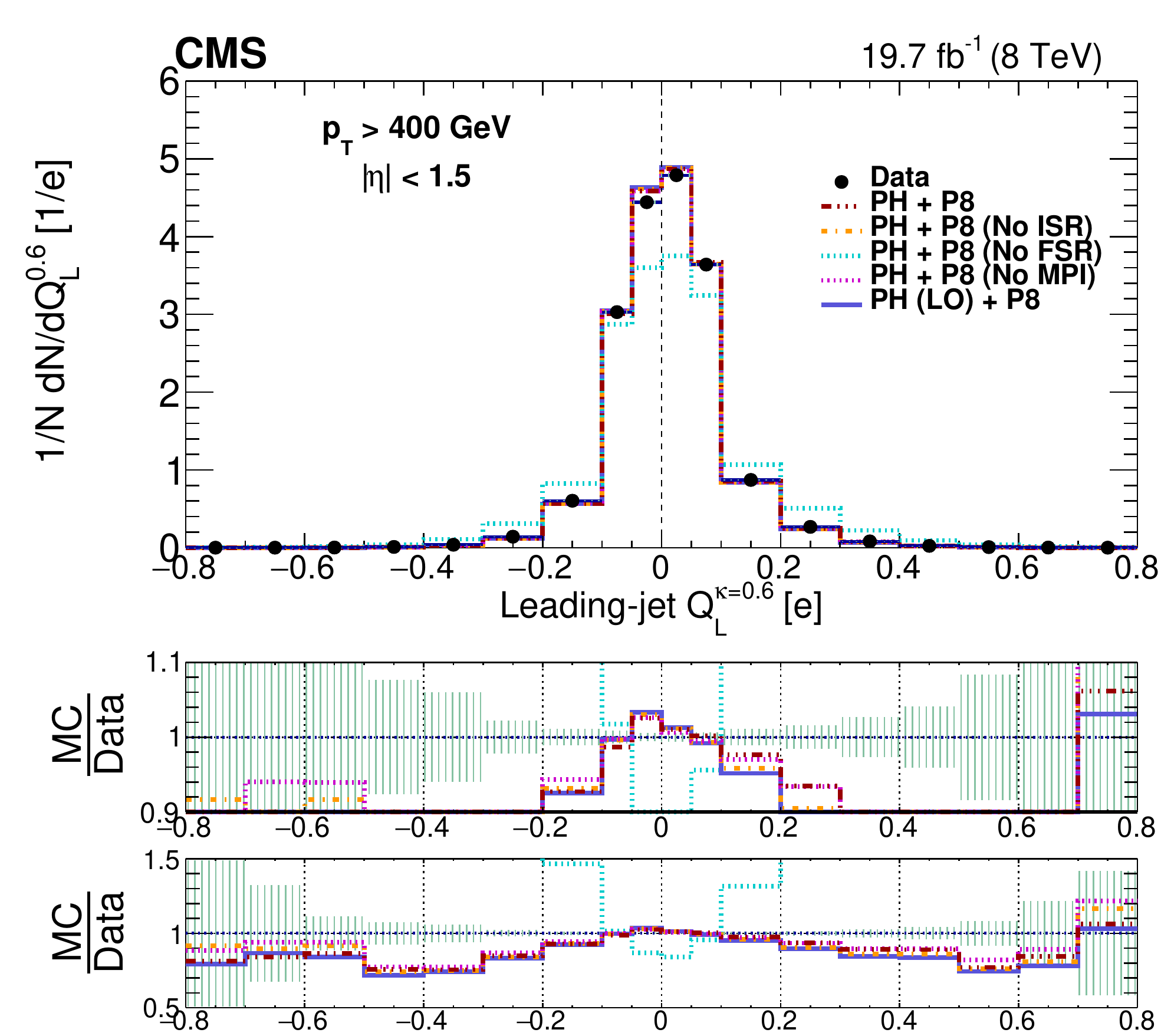}
\includegraphics[width=6.7cm]{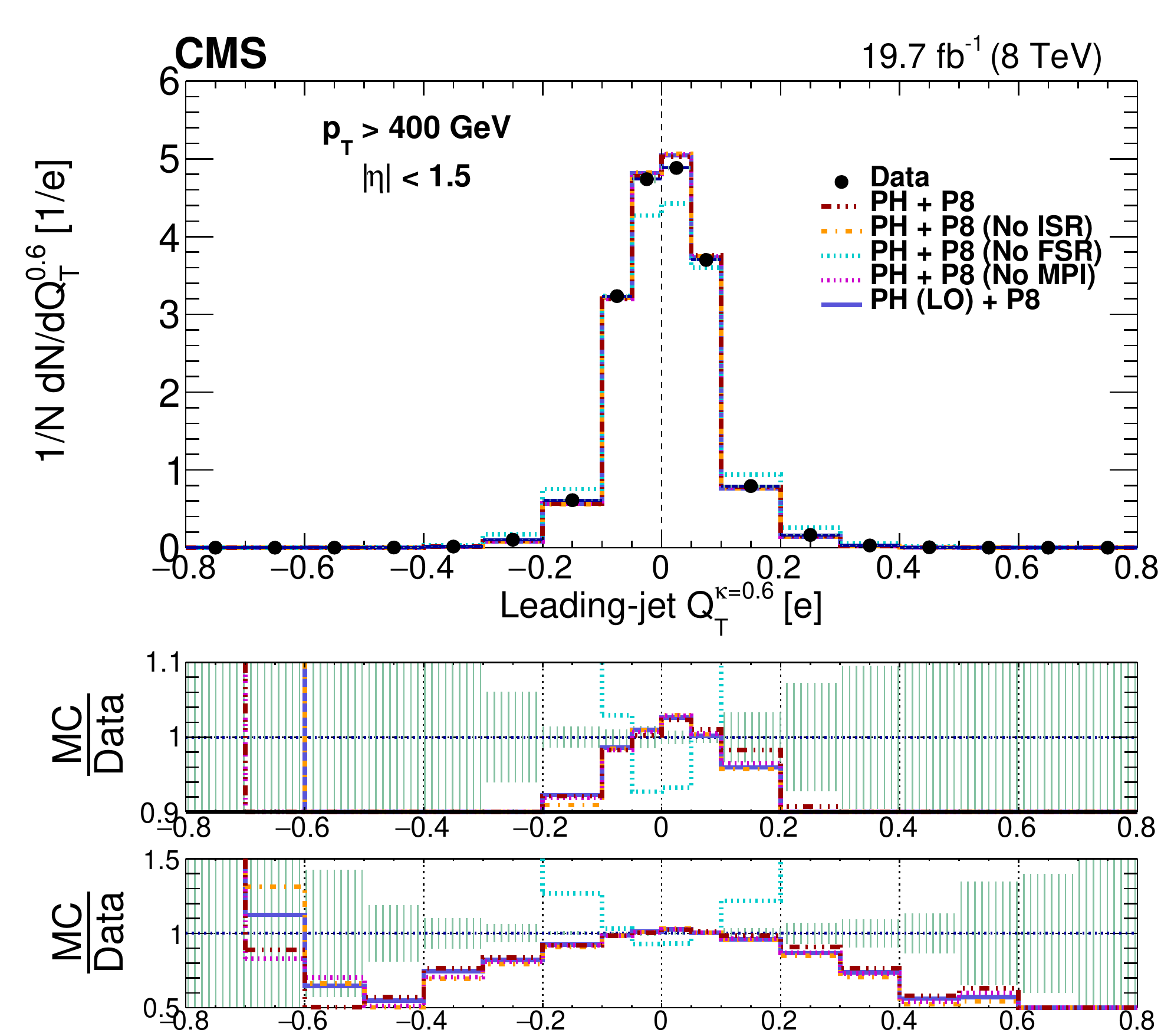}
\caption{\label{finalModel}Comparison of unfolded leading-jet charge distributions with predictions from \POWHEG+~\PYTHIA{8} (``PH+P8'').
The NLO \POWHEG prediction with the NLO CT10 PDF set is compared with predictions where initial-state radiation (``No ISR''), final-state radiation (``No FSR''), or multiple-parton interactions (``No MPI'') are disabled in \PYTHIA{8}. A LO \POWHEG prediction using the LO CTEQ6L1 PDF set (``LO'') is also shown.
The default jet charge definition ($Q^\kappa$), the longitudinal jet charge definition ($Q_{L}^\kappa$), and the transverse jet charge definition ($Q_{T}^\kappa$) are shown for $\kappa=0.6$.
Hashed uncertainty bands include both statistical and systematic contributions in data, added in quadrature.
The ratio of data to simulation is displayed twice below each plot with two different vertical scales.
}
\end{center}
\end{figure}

Figure~\ref{finalModel} presents the unfolded leading-\pt jet charge distributions for the three jet charge definitions introduced in Section~\ref{sec-definitions} with $\kappa=0.6$ compared to the generator level \POWHEG+~\PYTHIA{8} predictions for the CT10 NLO PDF set.
Each plot also displays the ratio of data to the MC prediction and a band representing the uncertainty determined by adding in quadrature the statistical uncertainties in the data and those arising from all systematic effects in the data. The distributions are normalized to unity.
The NLO \POWHEG predictions with the NLO CT10 PDF set are compared with predictions where initial-state radiation, final-state radiation, or multiple-parton interactions are disabled in \PYTHIA{8}.
They are also compared to a LO \POWHEG prediction that uses the LO CTEQ6L1 PDF set.
For all three jet charge definitions, the data is slightly broader than the prediction from \POWHEG+~\PYTHIA{8}.
The prediction for the jet charge distribution of the leading jet in the event is found to be rather insensitive to NLO QCD effects in the matrix-element calculation using \POWHEG since the jet charge distribution is changed by significantly less than the experimental uncertainty.
Similarly, simulations of initial-state radiation and multiple-parton interactions do not change the jet charge distribution.
Disabling the simulation of final-state radiation in \PYTHIA{8}, however, leads to a significantly broader jet charge distribution, from which it can be concluded that the jet charge distribution is mainly sensitive to the modeling of this effect.

\begin{figure}[p]
\begin{center}
\includegraphics[width=6.7cm]{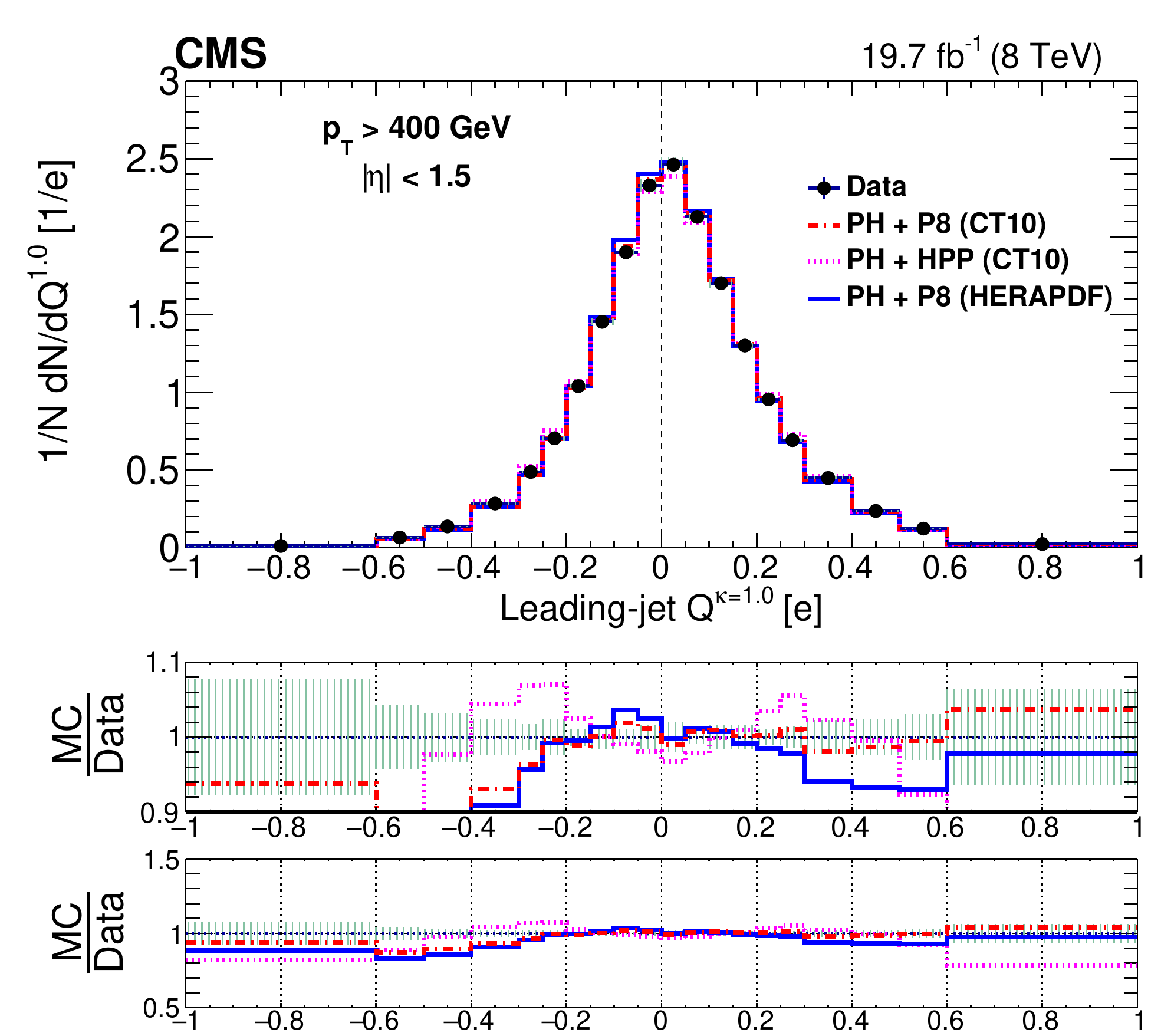}
\includegraphics[width=6.7cm]{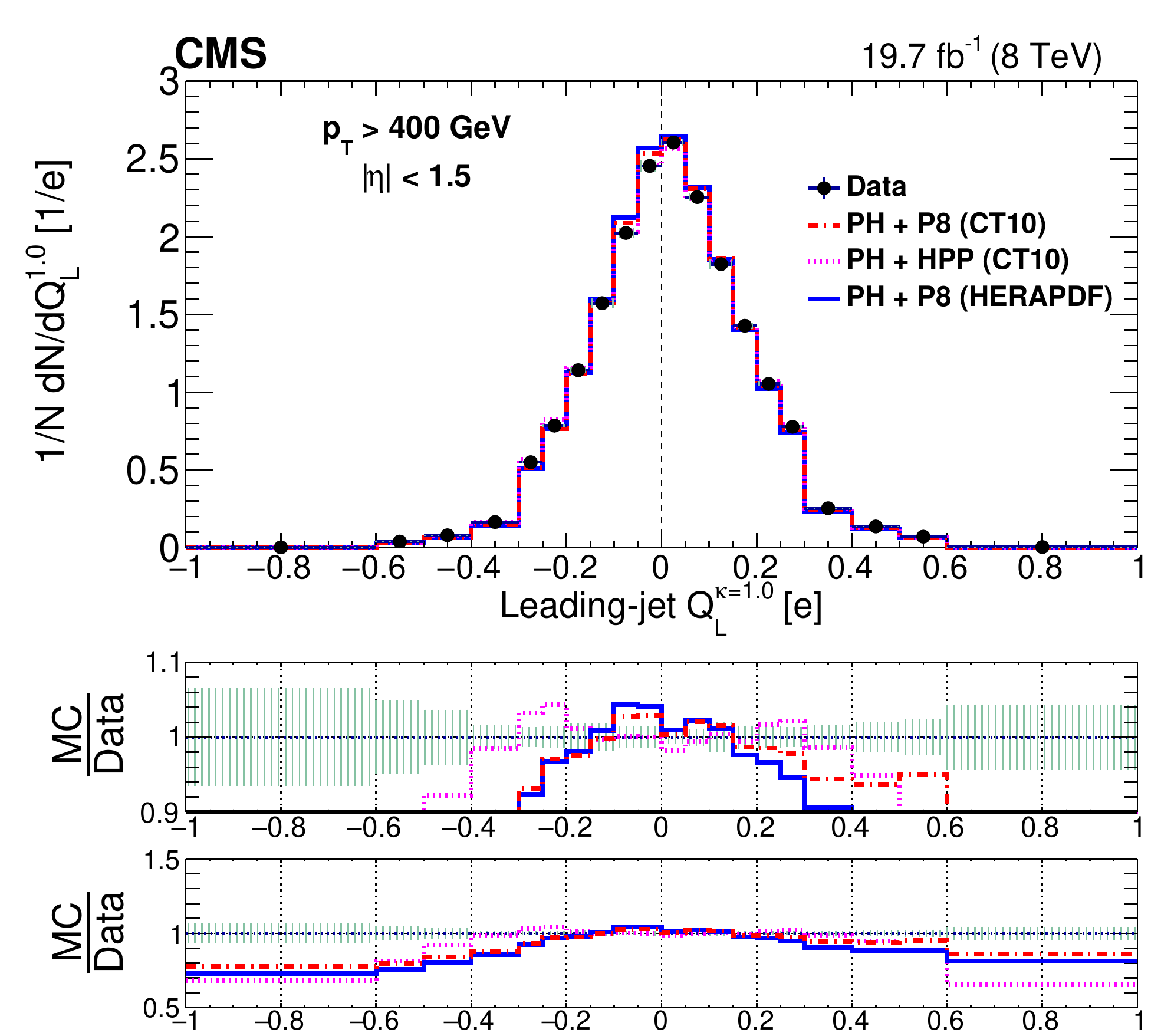} \\
\includegraphics[width=6.7cm]{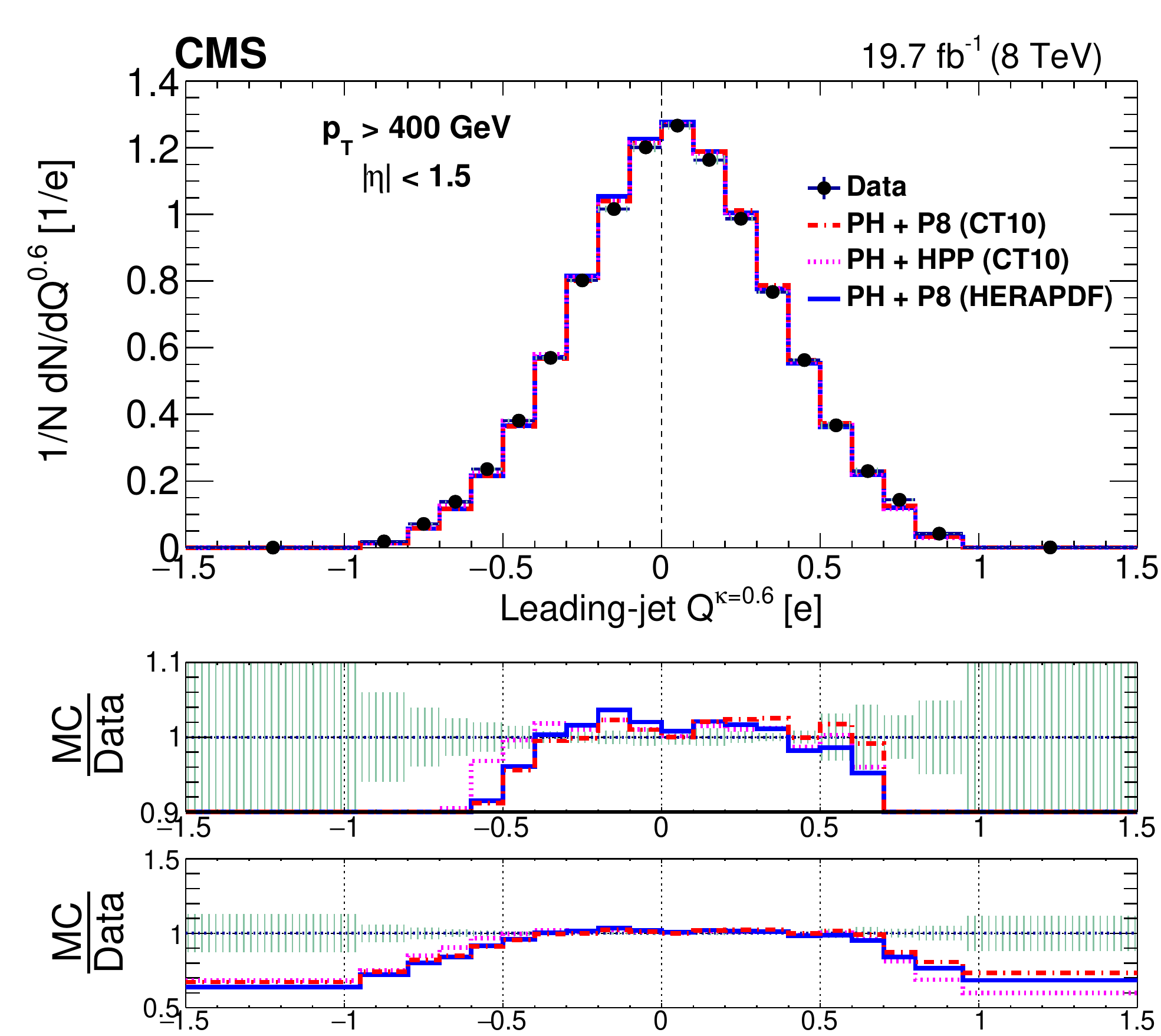}
\includegraphics[width=6.7cm]{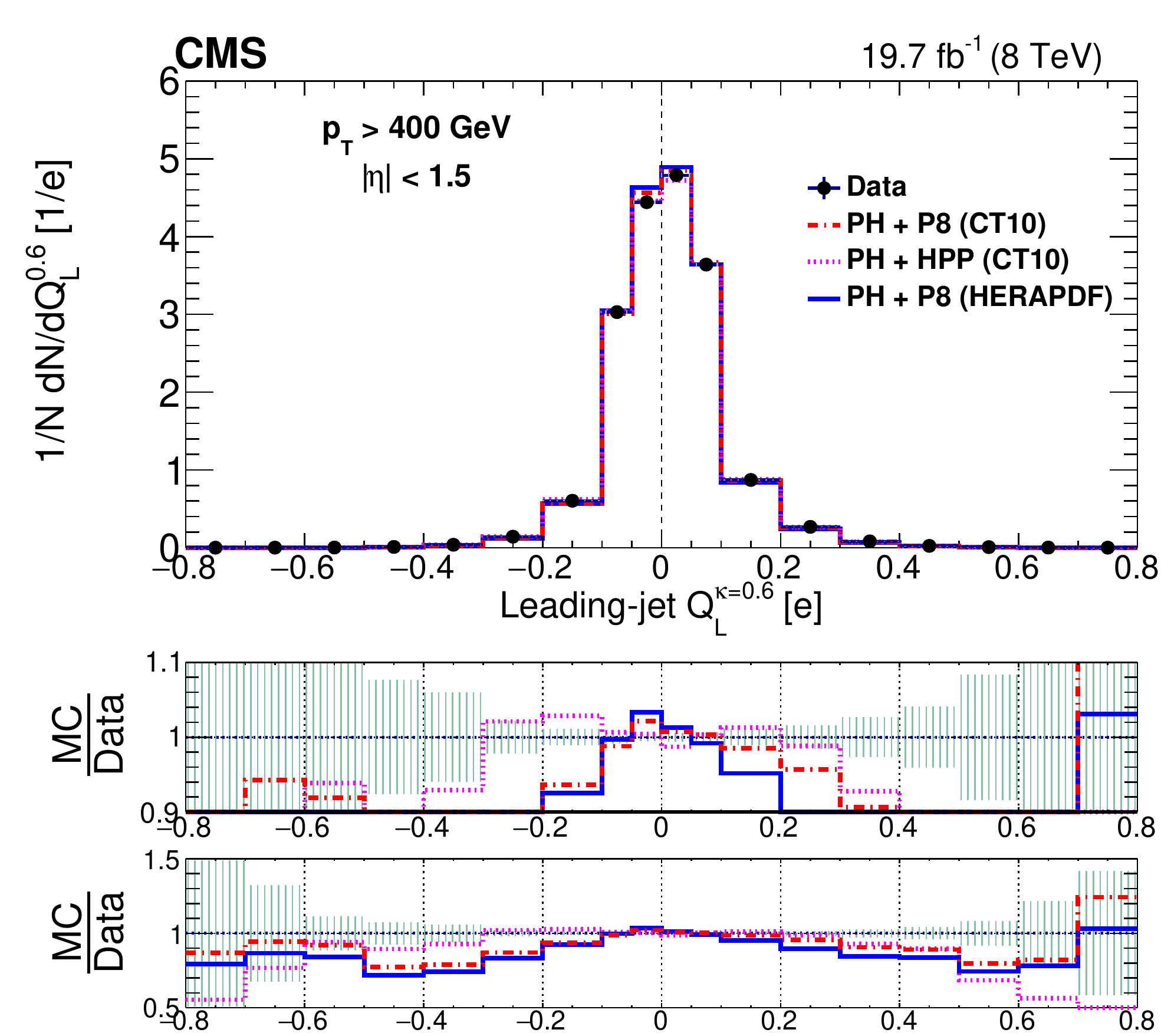} \\
\includegraphics[width=6.7cm]{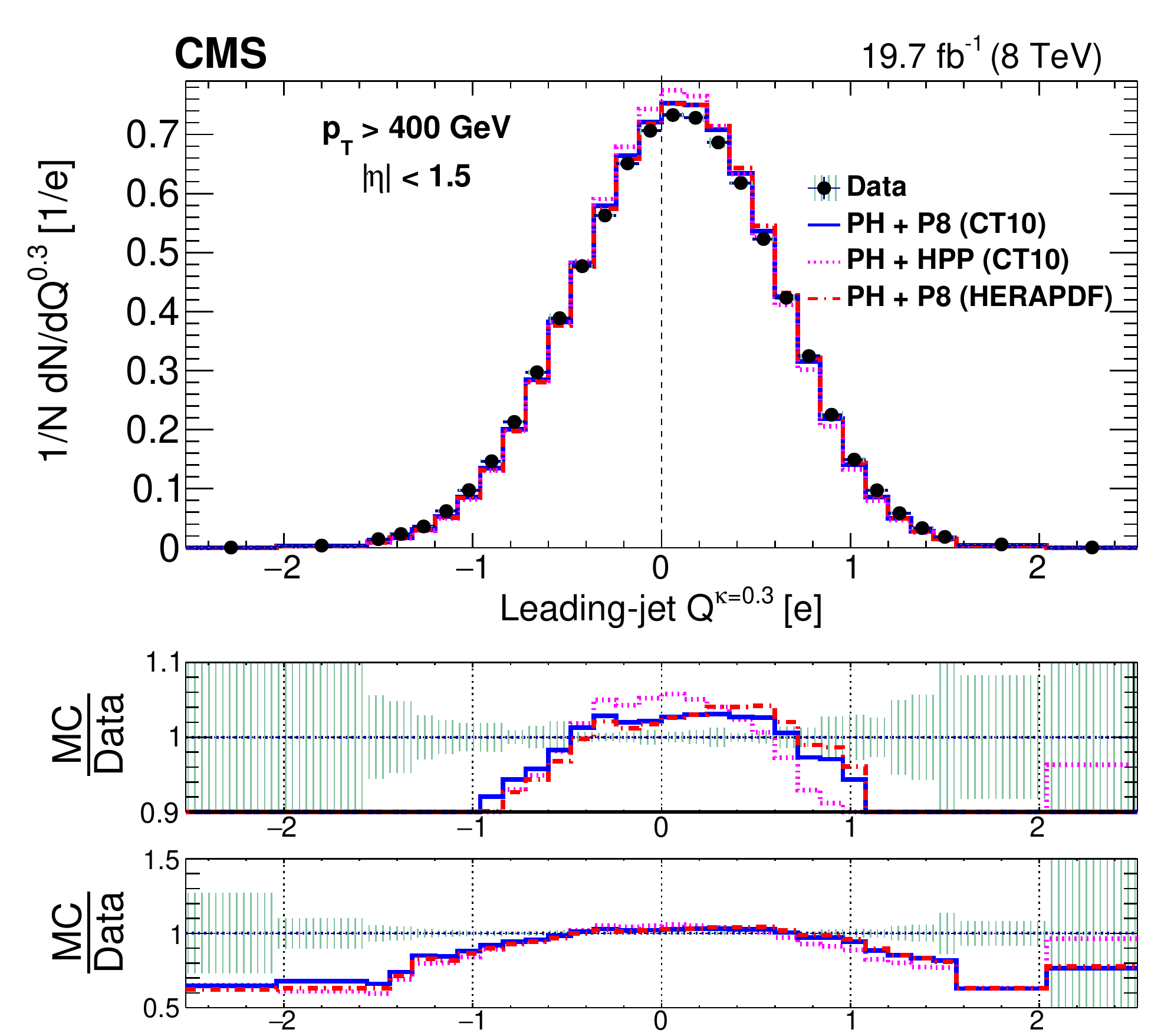}
\includegraphics[width=6.7cm]{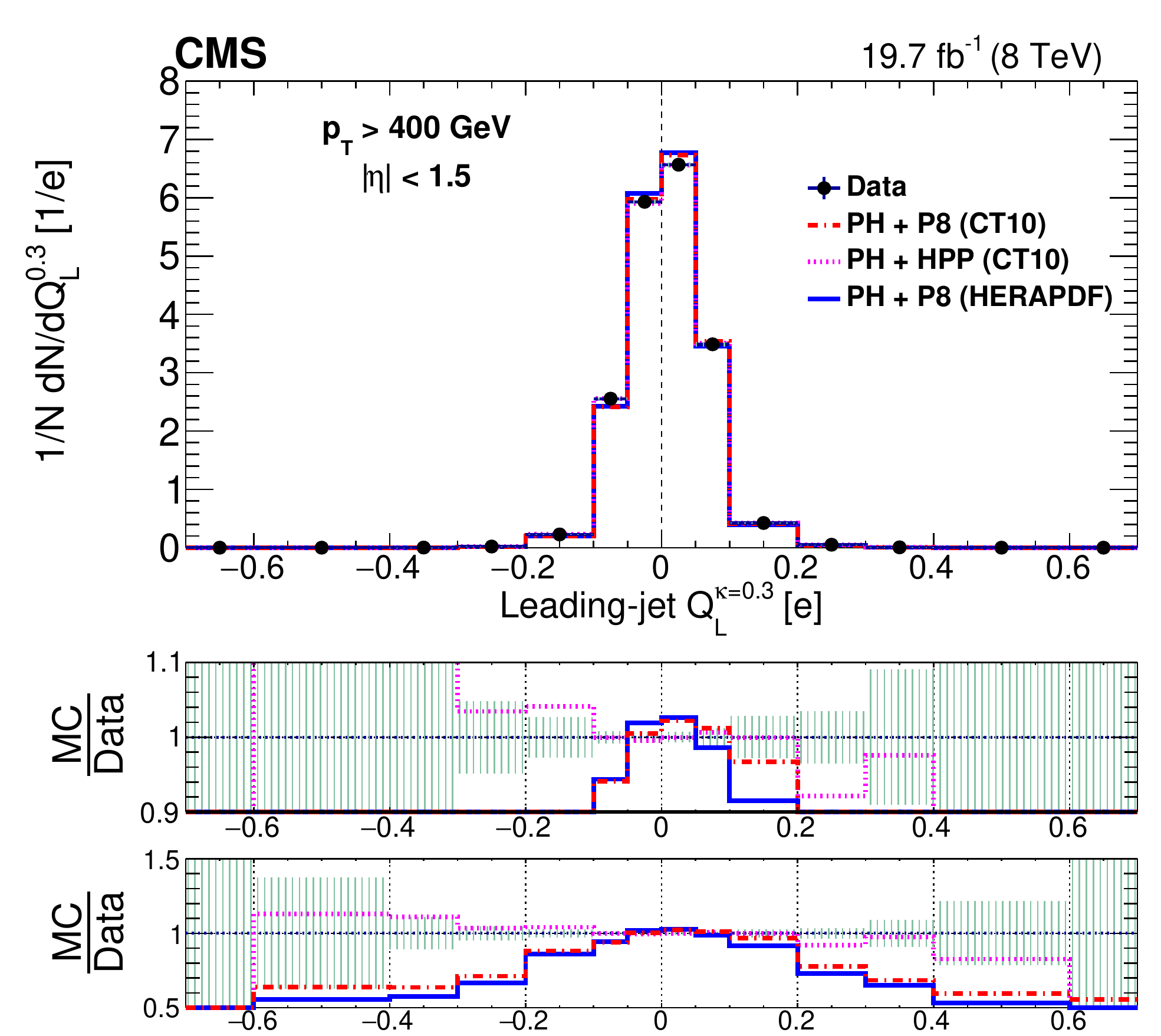}
\caption{\label{final}Comparison of unfolded leading-jet charge $Q^\kappa$ and $Q_{L}^\kappa$ distributions with \POWHEG+~\PYTHIA{8} (``PH+P8'') and \POWHEG+~\HERWIG{++} (``PH+HPP'') generators.
In addition to the \POWHEG+~\PYTHIA{8} predictions with the NLO CT10 PDF set (``CT10''), the distributions are also compared with the NLO HERAPDF~1.5 set (``HERAPDF'').
The left column shows the distributions for the default jet charge definition ($Q^\kappa$) with all three different $\kappa$ values, while the right column shows for the longitudinal jet charge definition ($Q_{L}^\kappa$) with all three different values of $\kappa$. Hashed uncertainty bands include both statistical and systematic contributions in data, added in quadrature.
The ratio of data to simulation is displayed twice below each plot with two different vertical scales.
}
\end{center}
\end{figure}

\begin{figure}[htb]
\begin{center}
\includegraphics[width=6.7cm]{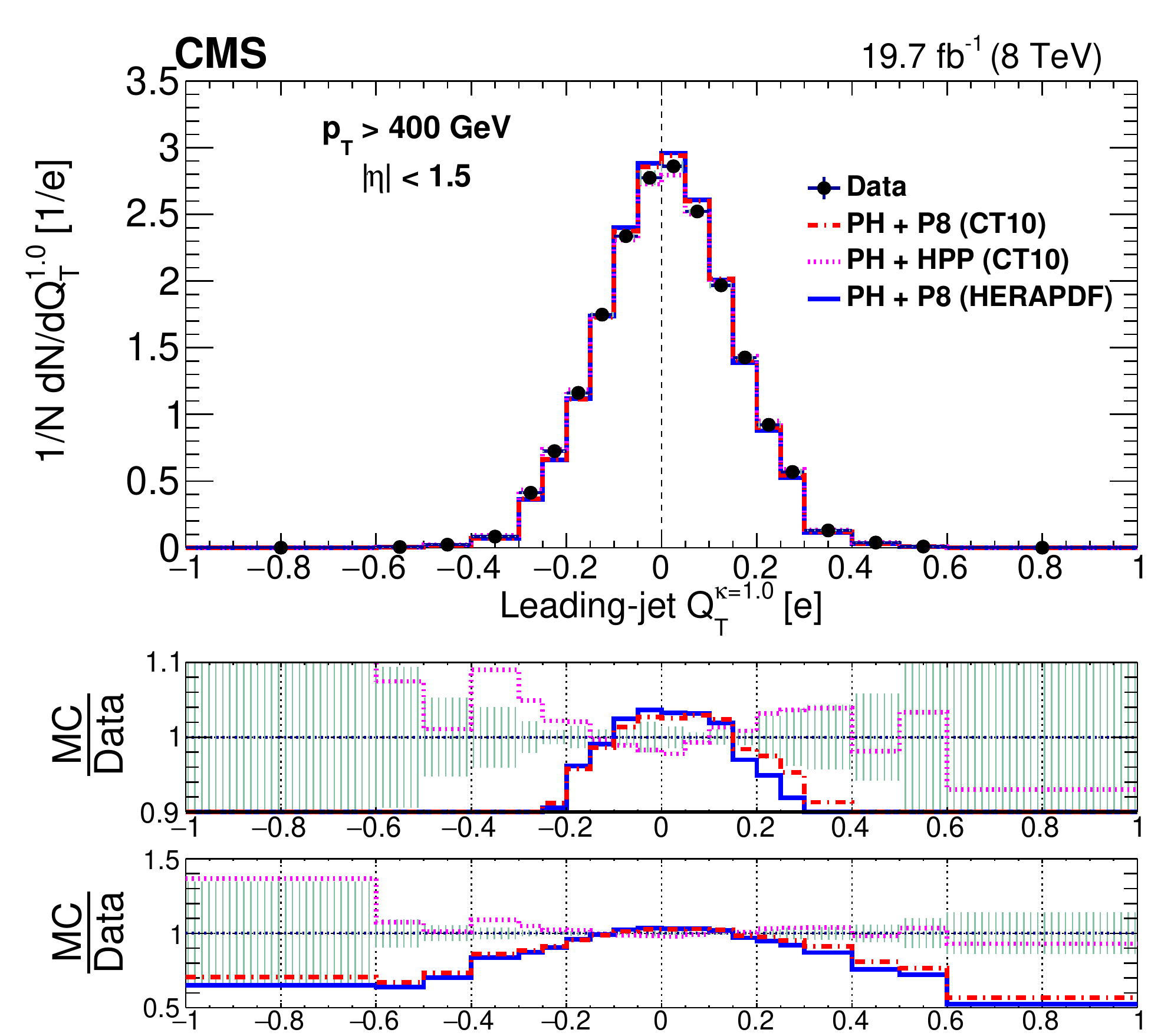}
\includegraphics[width=6.7cm]{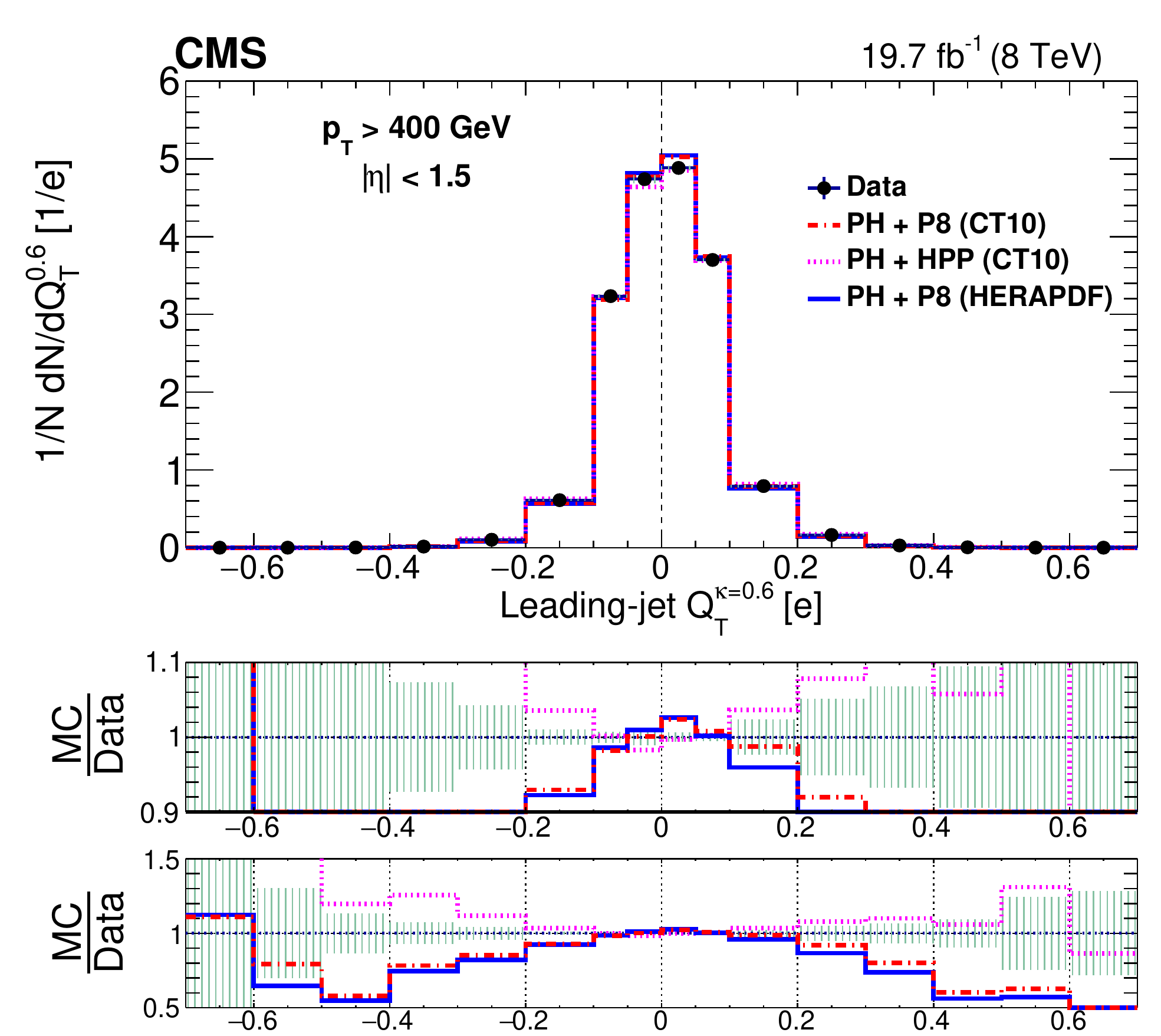} \\
\includegraphics[width=6.7cm]{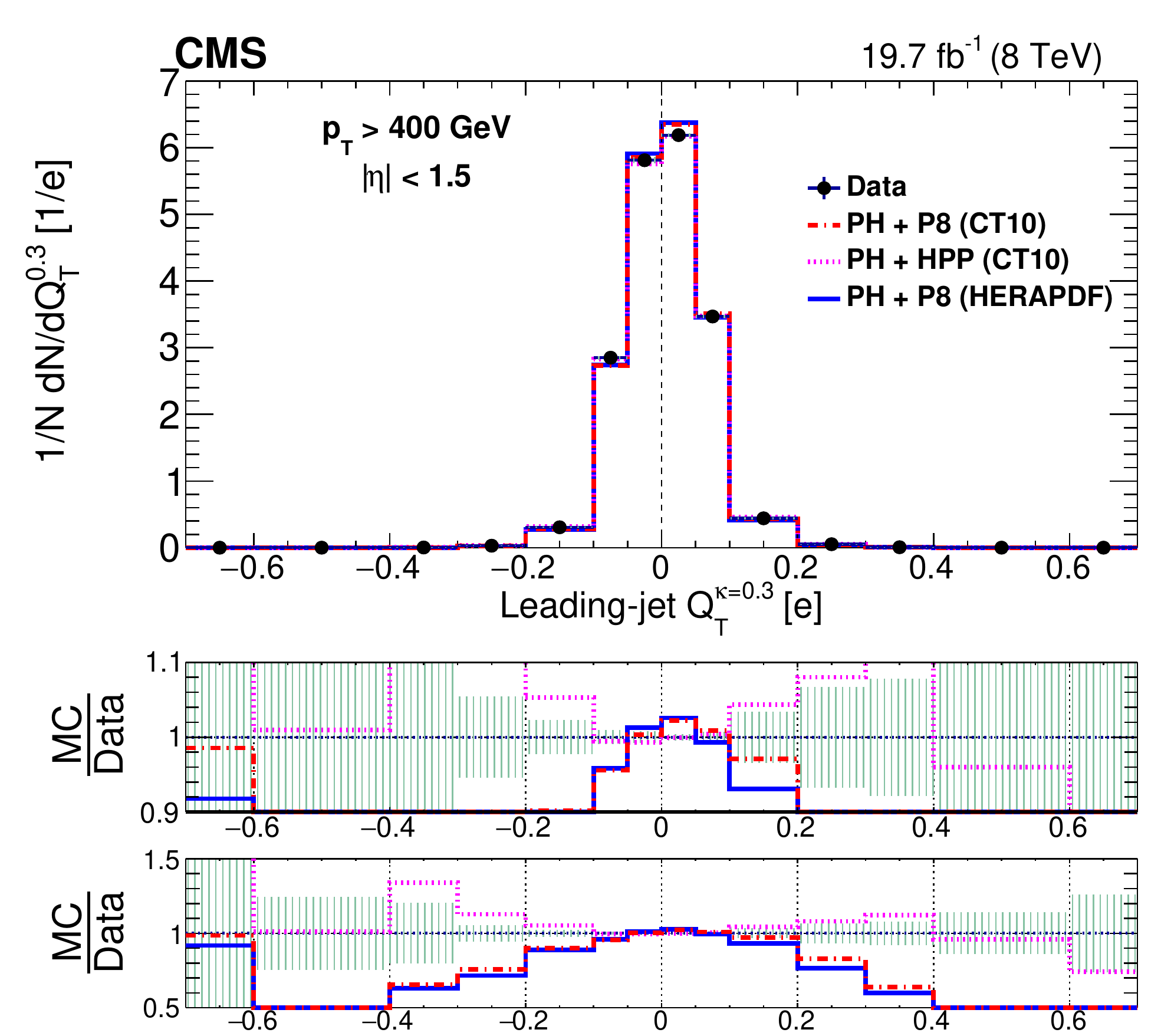}
\caption{\label{final1}Comparison of unfolded leading-jet charge distributions $Q_{T}^\kappa$ with \POWHEG+~\PYTHIA{8} (``PH+P8'') and \POWHEG+~\HERWIG{++} (``PH+HPP'') generators for transverse jet charge definition ($Q_{T}^\kappa$) with all different $\kappa$ values.
In addition to the \POWHEG+~\PYTHIA{8} predictions with the NLO CT10 PDF set (``CT10''), the distributions are also compared with the NLO HERAPDF~1.5 set (``HERAPDF'').
Hashed uncertainty bands include both statistical and systematic contributions in data, added in quadrature.
The ratio of data to simulation is displayed twice below each plot with two different vertical scales.
}
\end{center}
\end{figure}

\begin{figure}[p]
\begin{center}
\includegraphics[width=6.7cm]{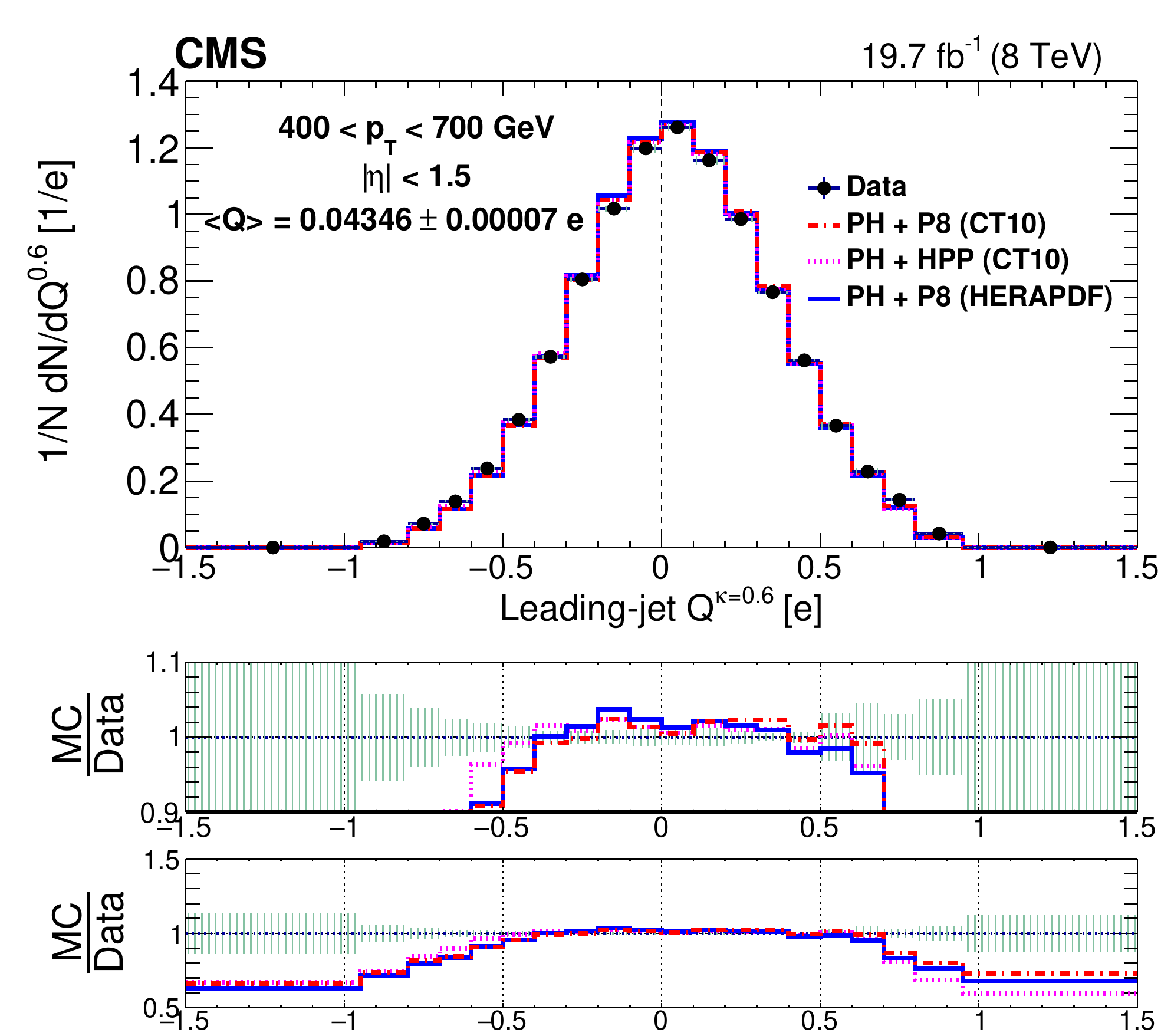}
\includegraphics[width=6.7cm]{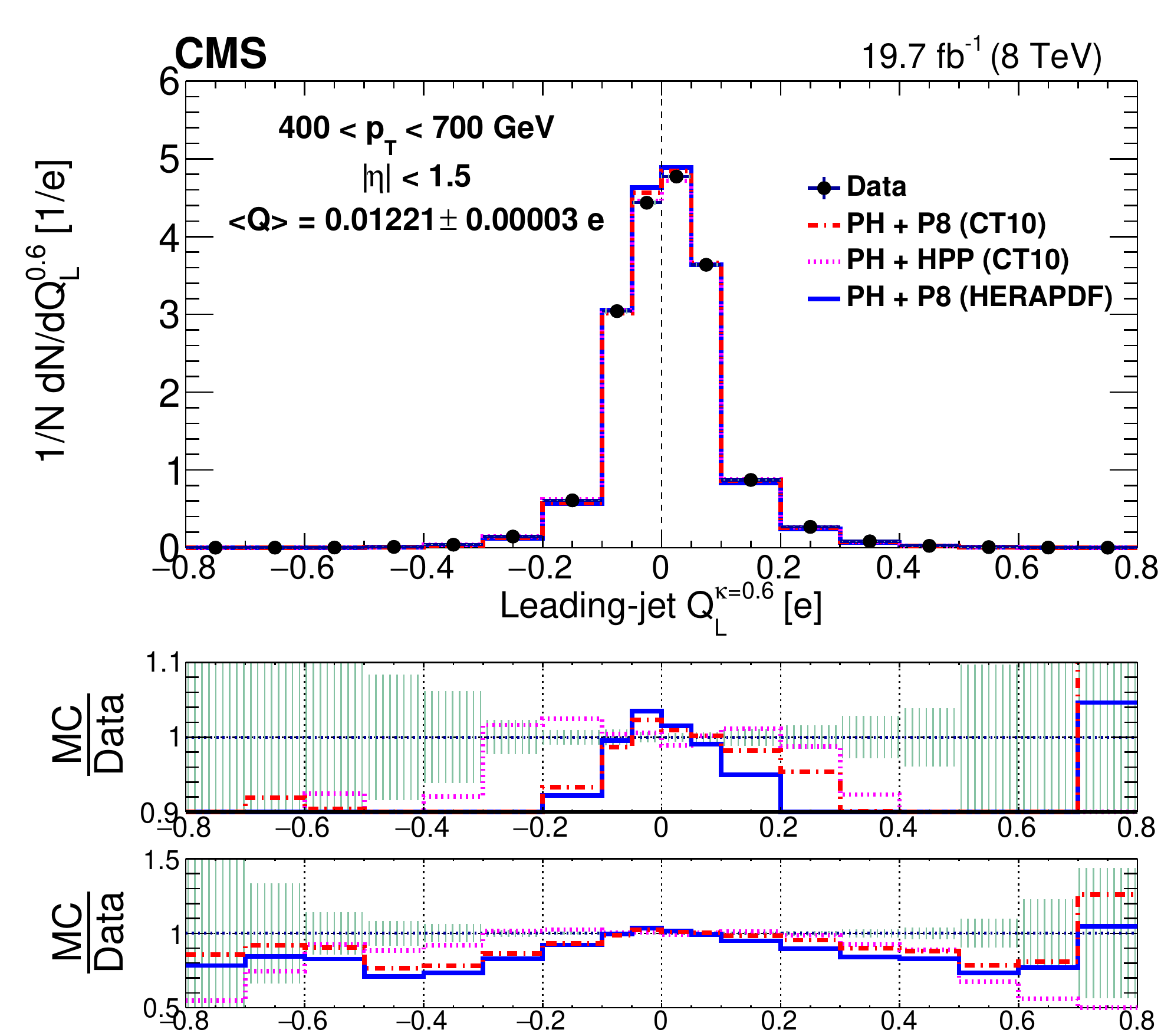} \\
\includegraphics[width=6.7cm]{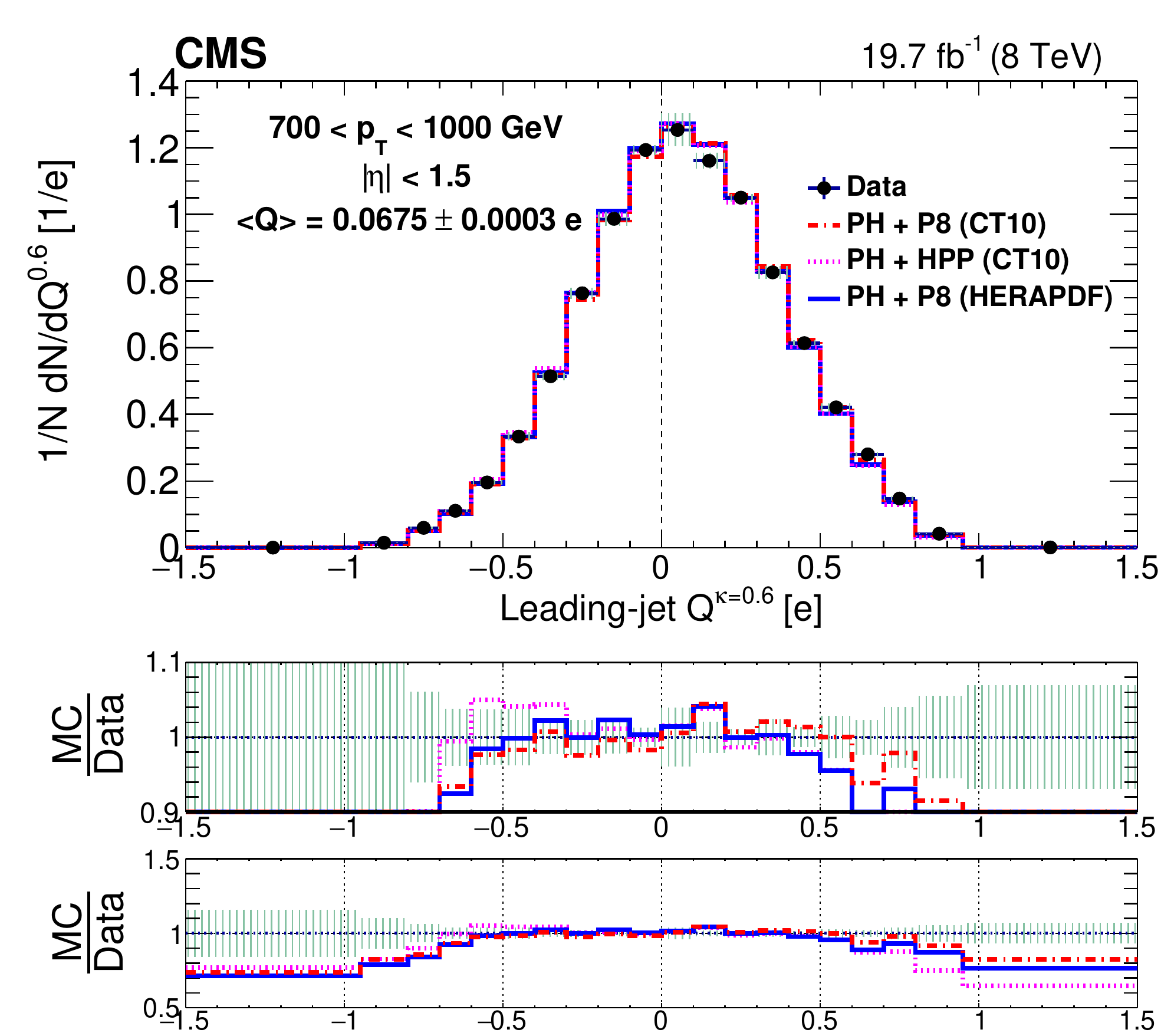}
\includegraphics[width=6.7cm]{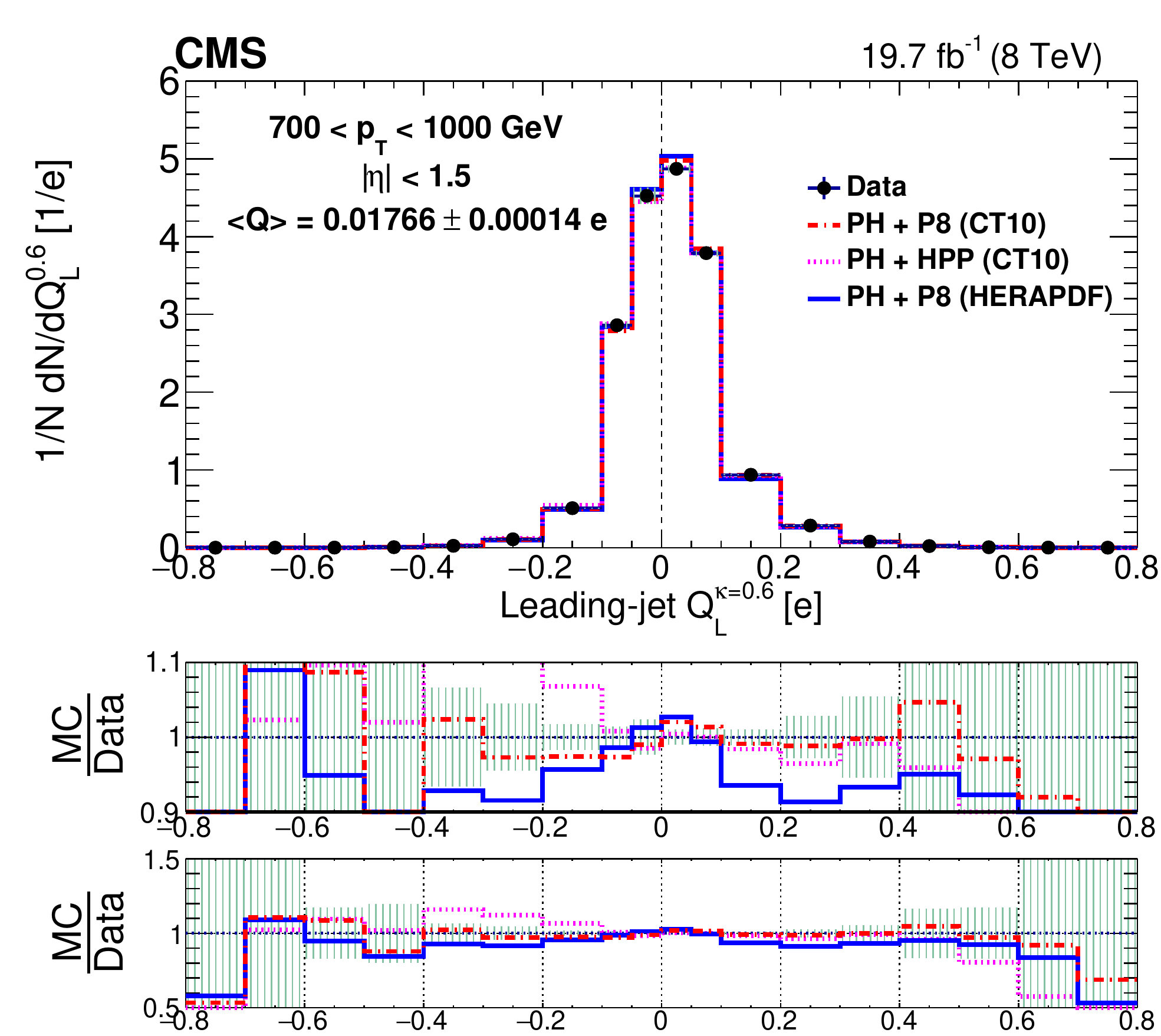} \\
\includegraphics[width=6.7cm]{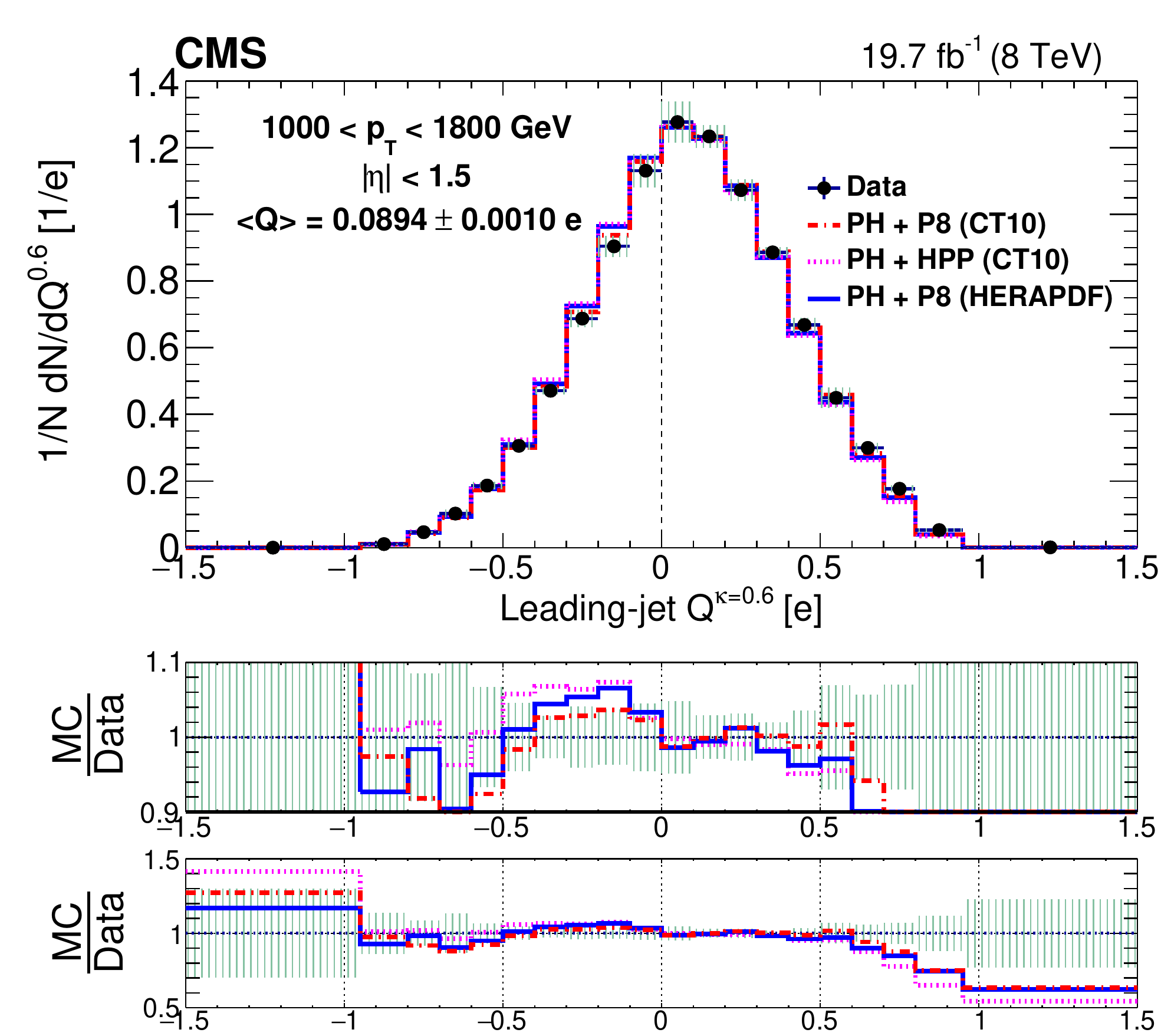}
\includegraphics[width=6.7cm]{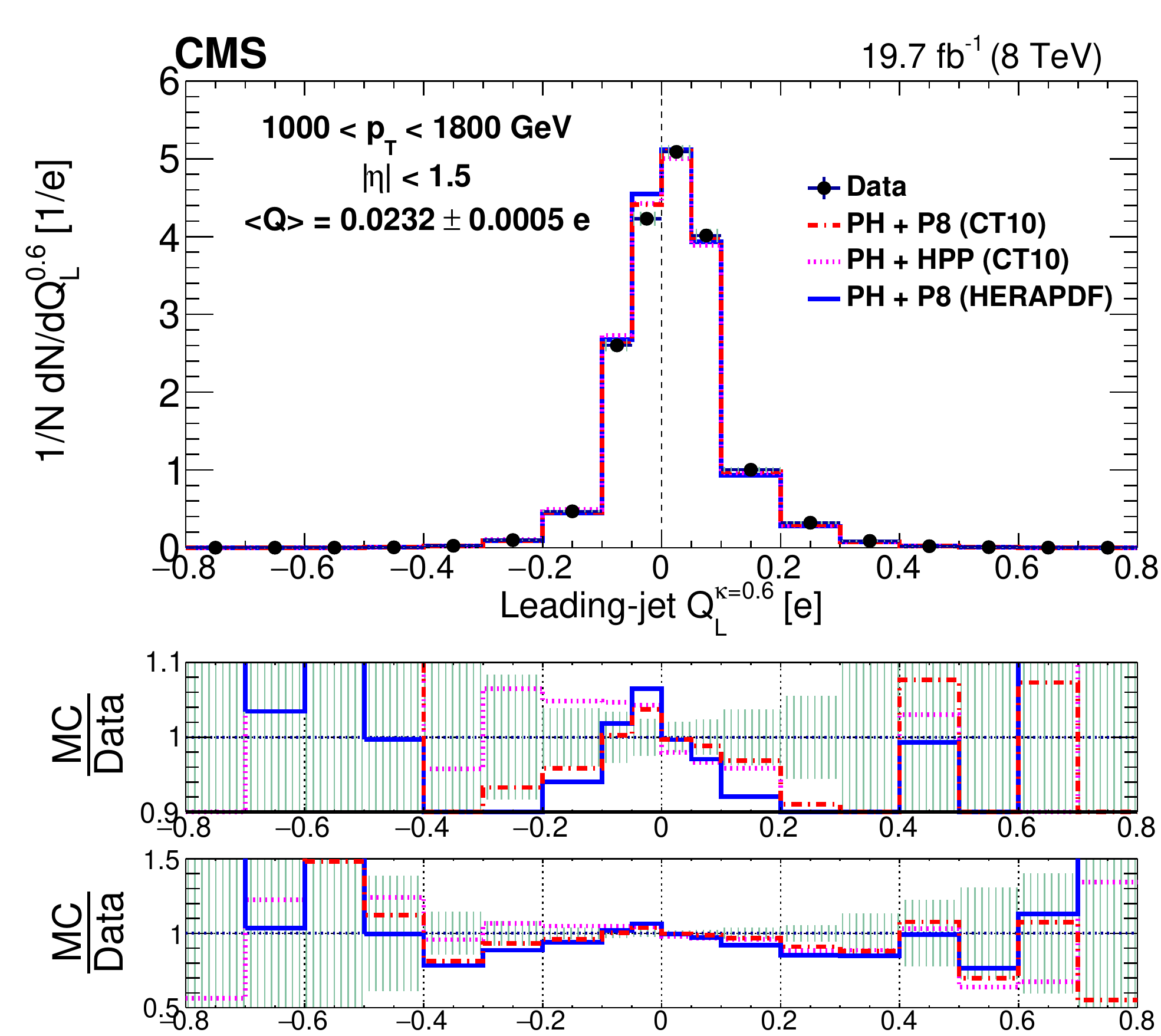}
\caption{\label{finalpt}Comparison of unfolded leading-jet charge distributions $Q^\kappa$ and $Q_{L}^\kappa$ with \POWHEG+~\PYTHIA{8} (``PH+P8'') and \POWHEG+~\HERWIG{++} (``PH+HPP'') generators in 3 ranges of leading-jet \pt.
In addition to the \POWHEG+~\PYTHIA{8} predictions with the NLO CT10 PDF set (``CT10''), the distributions are also compared with the NLO HERAPDF~1.5 set (``HERAPDF'').
The left column shows the jet \pt dependence for the default jet charge definition ($Q^\kappa$) with $\kappa$ = 0.6.
The right column shows the jet \pt dependence for the longitudinal jet charge definition ($Q_{L}^\kappa$) with $\kappa$ = 0.6. Hashed uncertainty bands include both statistical and systematic contributions in data, added in quadrature.
The ratio of data to simulation is displayed twice below each plot with two different vertical scales.
The average jet charge value is quoted on each panel only with statistical uncertainties.
}
\end{center}
\end{figure}

\begin{figure}[htb]
\begin{center}
\includegraphics[width=6.7cm]{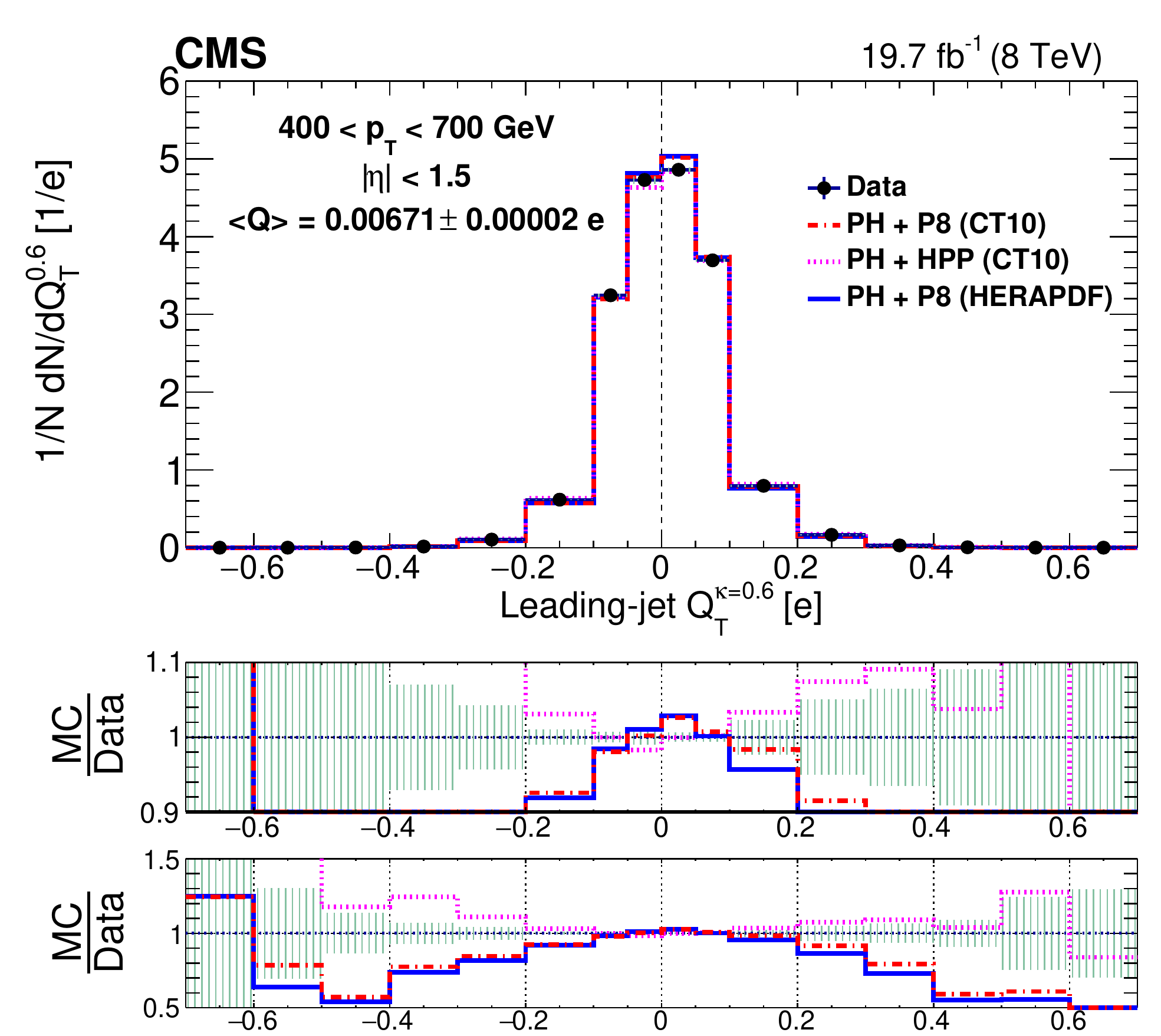}
\includegraphics[width=6.7cm]{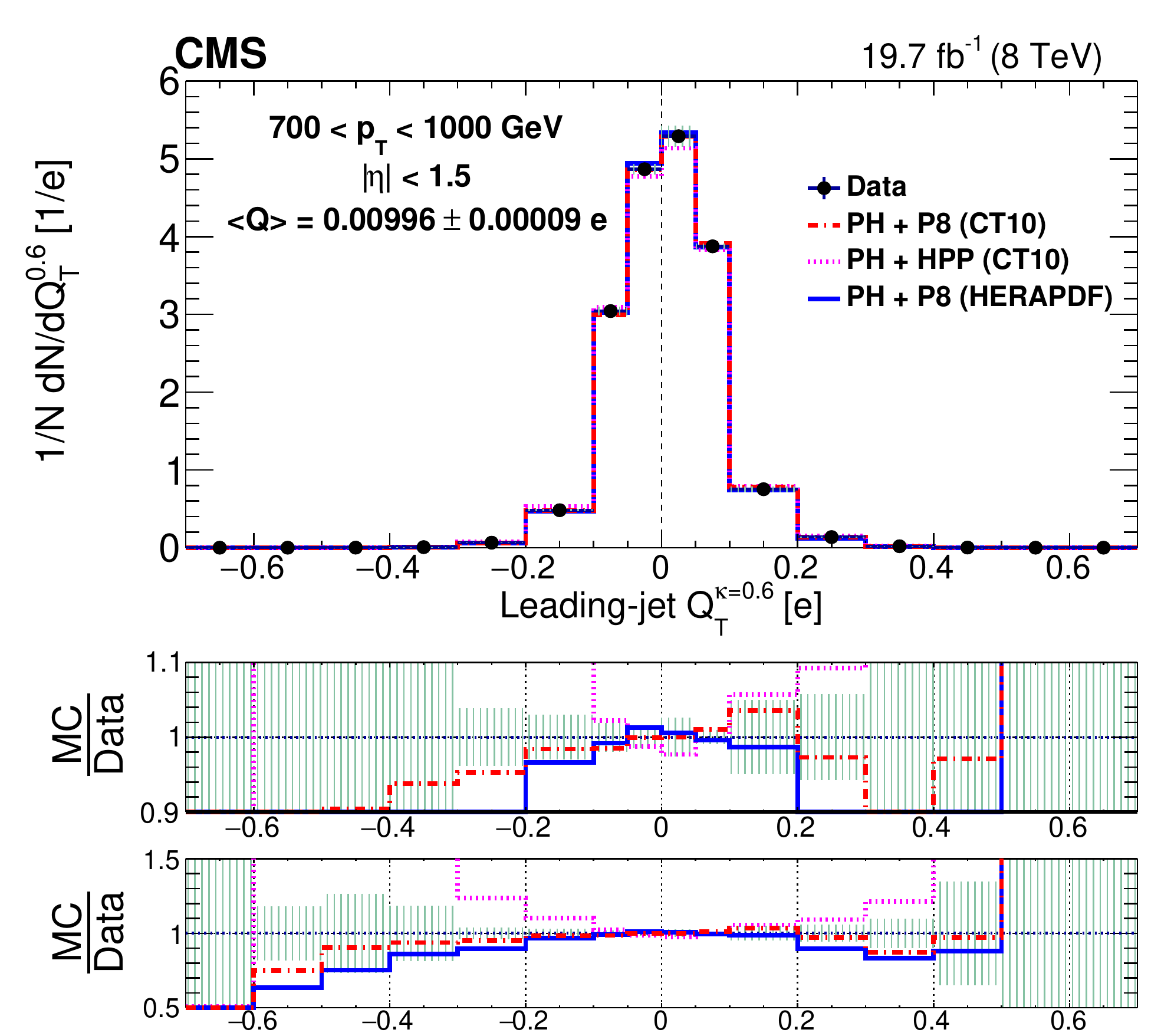} \\
\includegraphics[width=6.7cm]{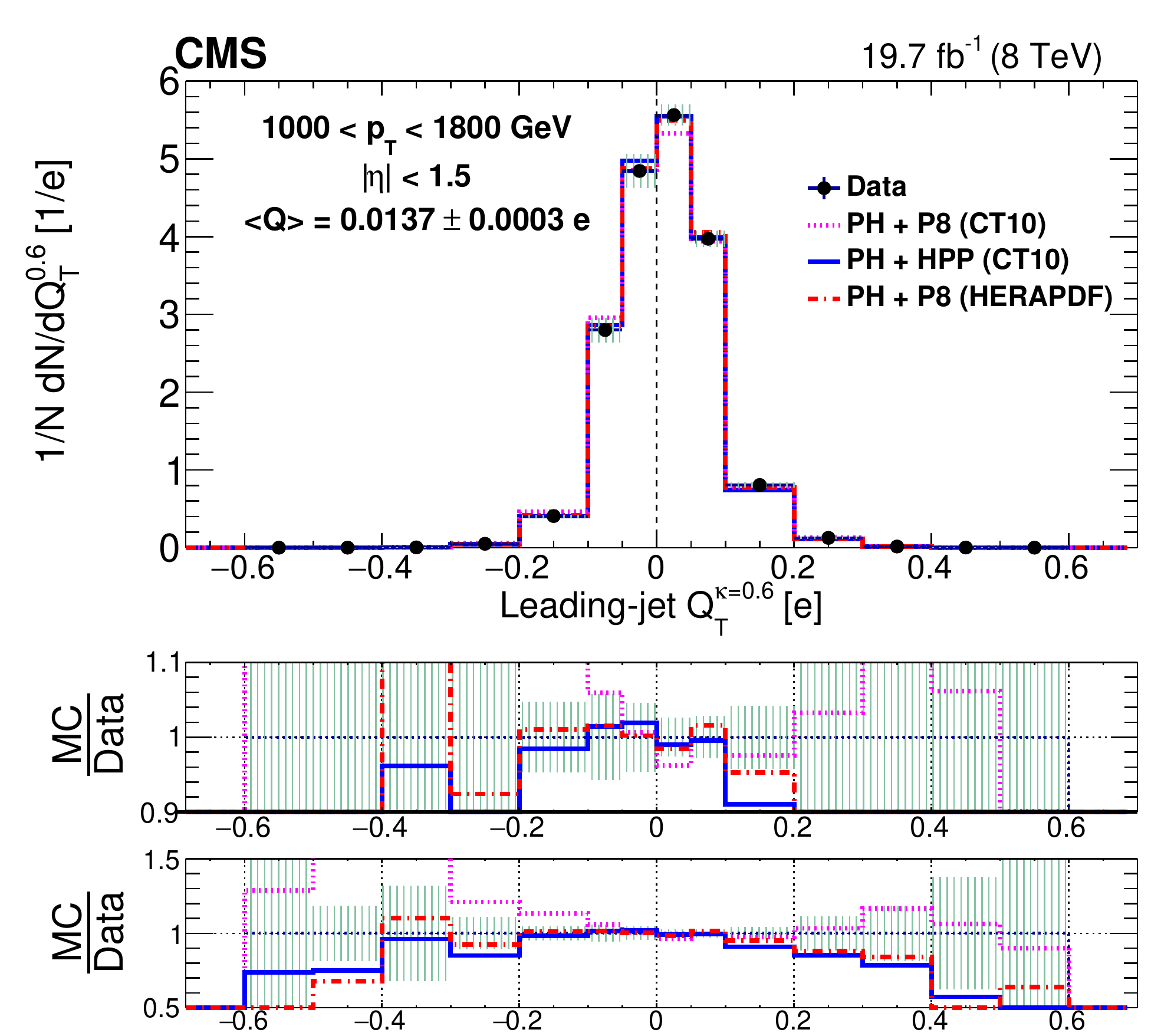}
\caption{\label{finalpt1}Comparison of unfolded leading-jet charge distributions $Q_{T}^\kappa$ with \POWHEG+~\PYTHIA{8} (``PH+P8'') and \POWHEG+~\HERWIG{++} (``PH+HPP'') generators in 3 ranges of leading-jet \pt for the transverse jet charge definition ($Q_{T}^\kappa$) with $\kappa$ = 0.6.
In addition to the \POWHEG+~\PYTHIA{8} predictions with the NLO CT10 PDF set (``CT10''), the distributions are also compared with the NLO HERAPDF~1.5 set (``HERAPDF'').
Hashed uncertainty bands include both statistical and systematic contributions in data, added in quadrature.
The ratio of data to simulation is displayed twice below each plot with two different vertical scales.
The average jet charge value is quoted on each panel only with statistical uncertainties.
}
\end{center}
\end{figure}

Figures~\ref{final}--\ref{finalpt1} present the distributions of the unfolded data compared to the generator-level \POWHEG+~\PYTHIA{8} and \POWHEG+~\HERWIG{++} predictions using the CT10 and HERAPDF~1.5 NLO PDF sets with \POWHEG+~\PYTHIA{8}.
The effect of the PS and fragmentation model on the jet charge distribution can be seen by comparing the predictions from \POWHEG+~\PYTHIA{8} with \POWHEG+~\HERWIG{++} simulations, which make predictions based on different models of parton showering and fragmentation. The effect of the PDF set on the jet charge distribution can be seen by comparing predictions with CT10 and HERAPDF~1.5.
For this comparison, CT10 is chosen as a widely used general PDF set, while HERAPDF~1.5 represents an alternative that shows differences of order 10\% in the predicted inclusive jet cross section~\cite{SMP-14-001} that are still compatible with the measurements in the region of interest, $\pt > 400\GeV$.

The dependence of the default and the longitudinal jet charge on different $\kappa$ values is demonstrated in Fig.~\ref{final}, while that for the transverse definition is given in Fig.~\ref{final1}.
The differences between \POWHEG+~\PYTHIA{8} and \POWHEG+~\HERWIG{++} in each jet charge can be quantified by the measure defined in Eq.~(\ref{errordef}). While for $Q_{T}^{0.6}$ and $Q_{L}^{0.6}$ it is found to be 2.5 and 2.6\% respectively, it is only 1.2\% for $Q^{0.6}$, showing a different sensitivity of the variables to the showering and fragmentation models.
The difference between predictions using CT10 and HERAPDF~1.5 PDF sets is found to be significantly smaller. 
Thus, the knowledge of the quark and gluon composition of the dijet sample defined by the PDF set is somewhat better than the knowledge of the parton shower and fragmentation modeling for the jet charge.

In general, the predictions from the \POWHEG+~\PYTHIA{8} and \POWHEG+~\HERWIG{++} generators show only mild discrepancies with data, although certain systematic differences are apparent. Experimental uncertainties are generally larger for small values of $\kappa$ as well as for $Q_{T}^\kappa$ because of the larger weights given to soft particles. For the $Q^\kappa$ and $Q_{L}^\kappa$ shown in Fig.~\ref{final}, \POWHEG+~\PYTHIA{8} and \POWHEG+~\HERWIG{++} show similar levels of agreement. For the $Q_{T}^\kappa$ given in Fig.~\ref{final1}, both generators diverge significantly from data in most of the range. The two generators differ systematically for the three definitions of jet charge, and we conclude that this measurement can constrain such modeling predictions. It should also be recognized that a smaller fraction of the differences between data and the simulation may arise from the choice of the PDF set, while a larger fraction of the differences may arise from assumptions about hadronization and parton showering.

Figure~\ref{finalpt} gives the dependence of the default and longitudinal jet charge on jet \pt.
The dependence of the transverse charge is shown in Fig.~\ref{finalpt1}. In the \pt range considered, the gluon fraction is expected to decrease with \pt from about 35\% in top panels to 15\% in the lower panels.
In general for all jet charge definitions, the level of agreement between the two generators increases as a function of jet \pt. This suggests that the description of gluon jets differs more between \POWHEG+~\PYTHIA{8} and \POWHEG+~\HERWIG{++} than the description of quark jets. The level of agreement between simulation and data remains similar as a function of jet \pt, while the \POWHEG+~\PYTHIA{8} and \POWHEG+~\HERWIG{++} predictions approach each other at large \pt.

\begin{figure}[p]
\begin{center}
\includegraphics[width=6.7cm]{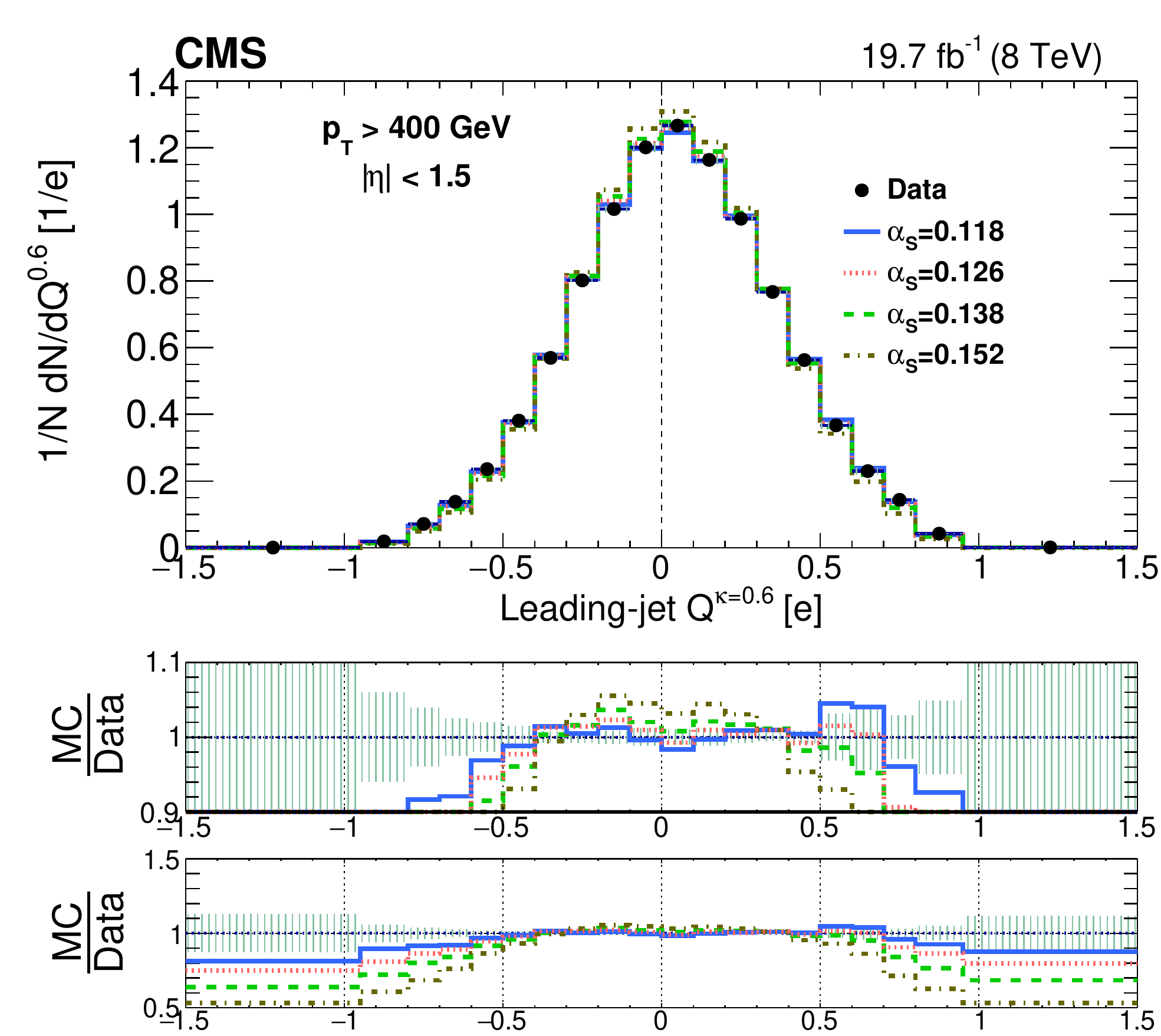}
\includegraphics[width=6.7cm]{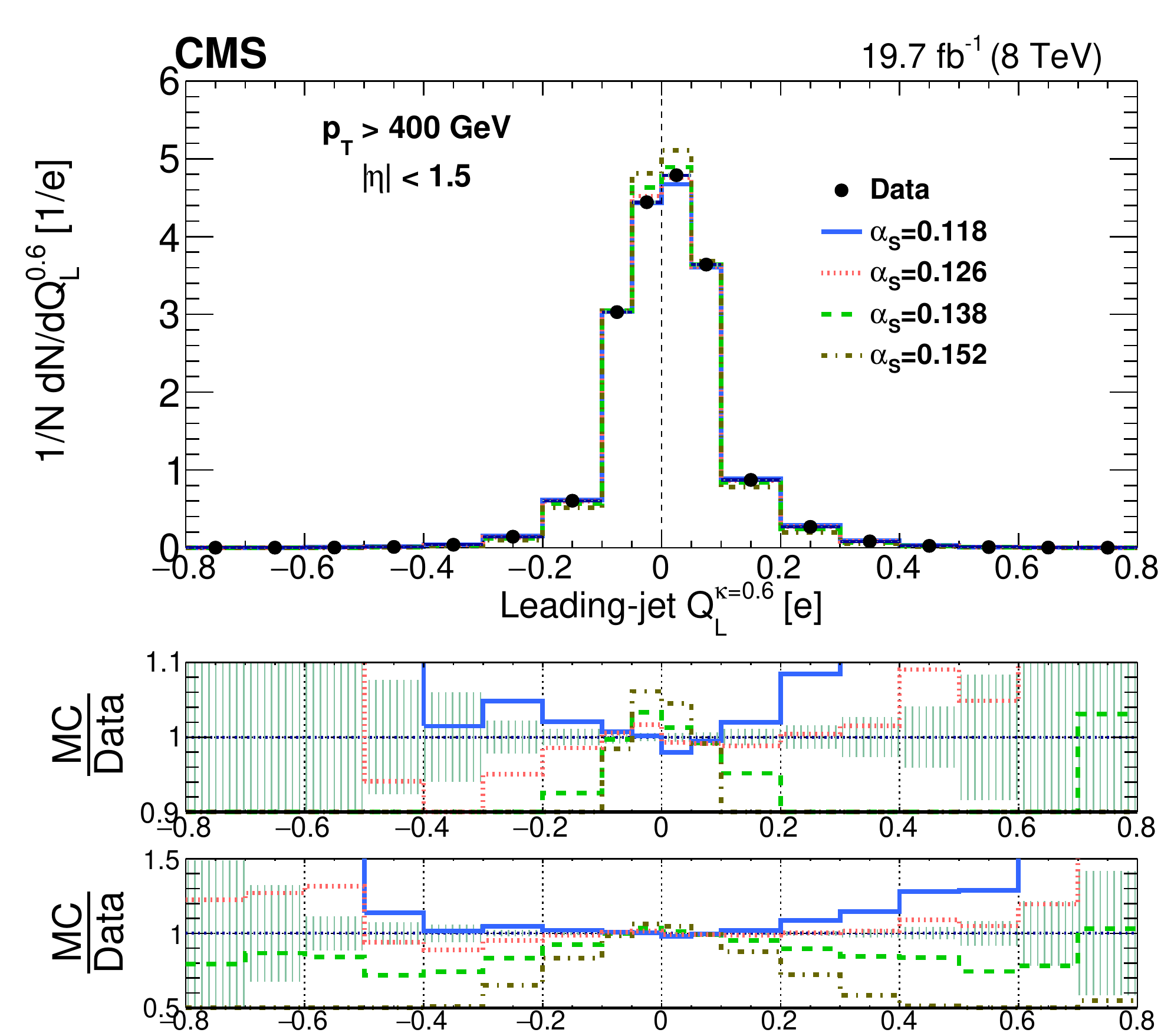} \\
\includegraphics[width=6.7cm]{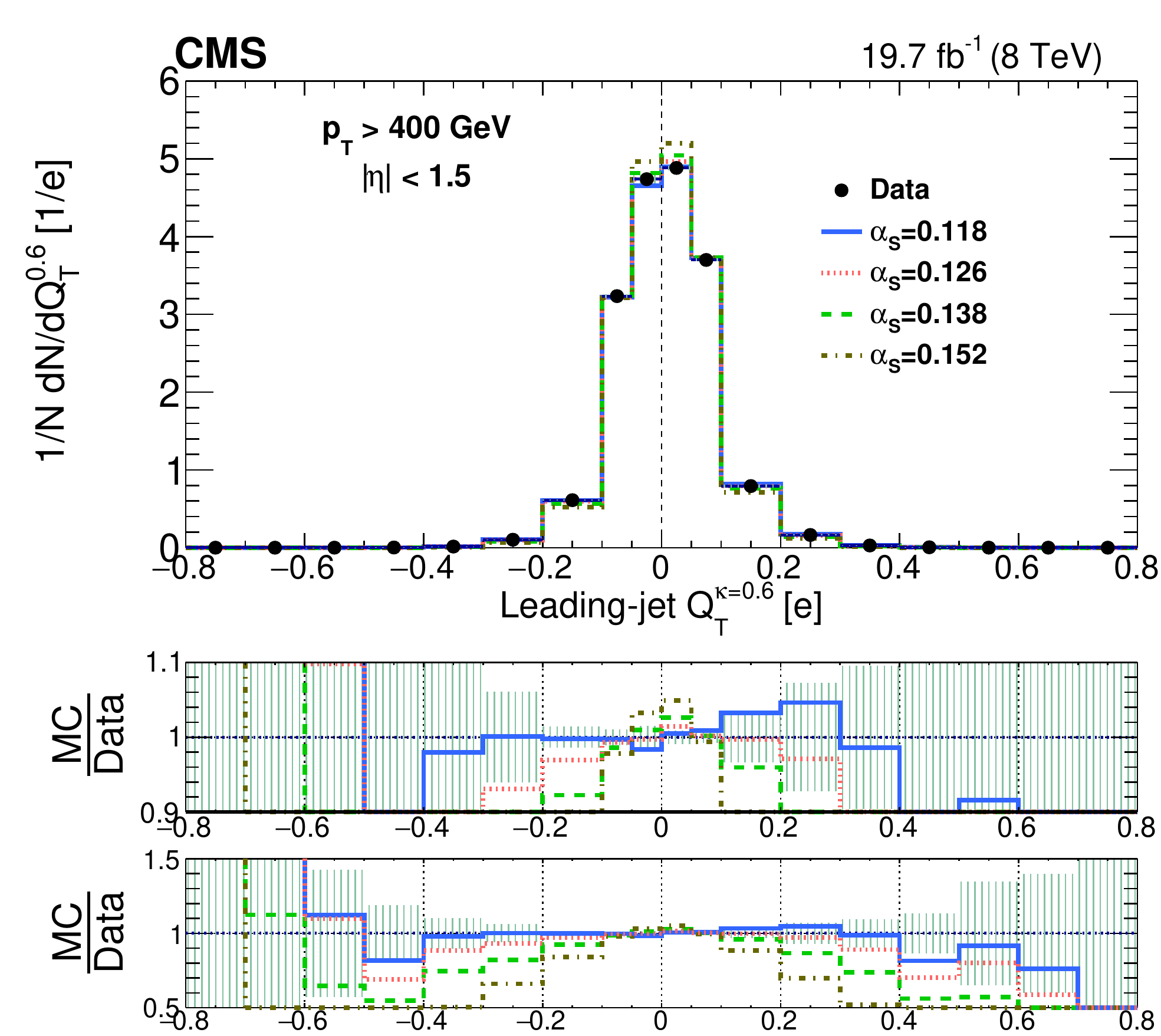} \\
\includegraphics[width=6.7cm]{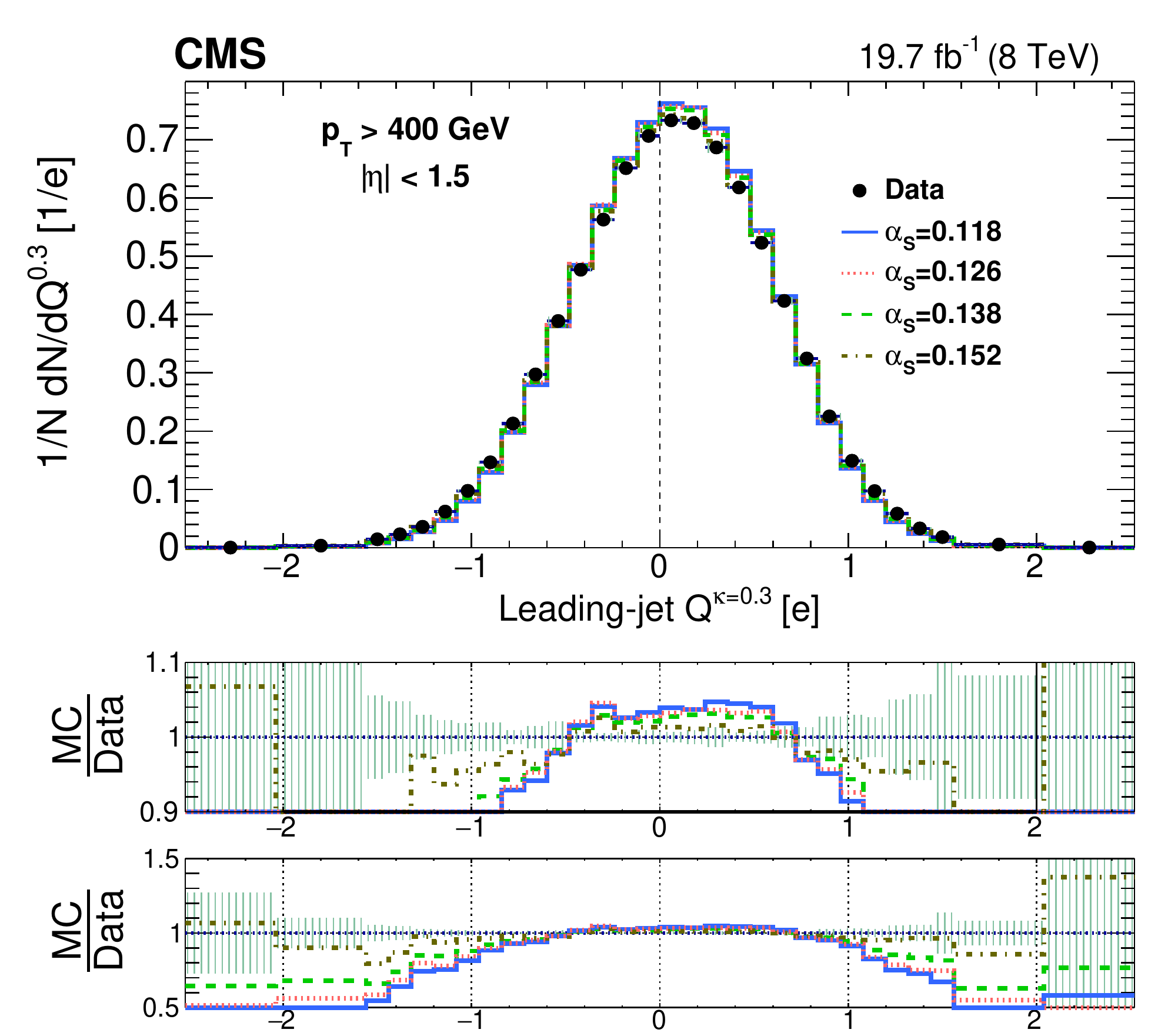}
\includegraphics[width=6.7cm]{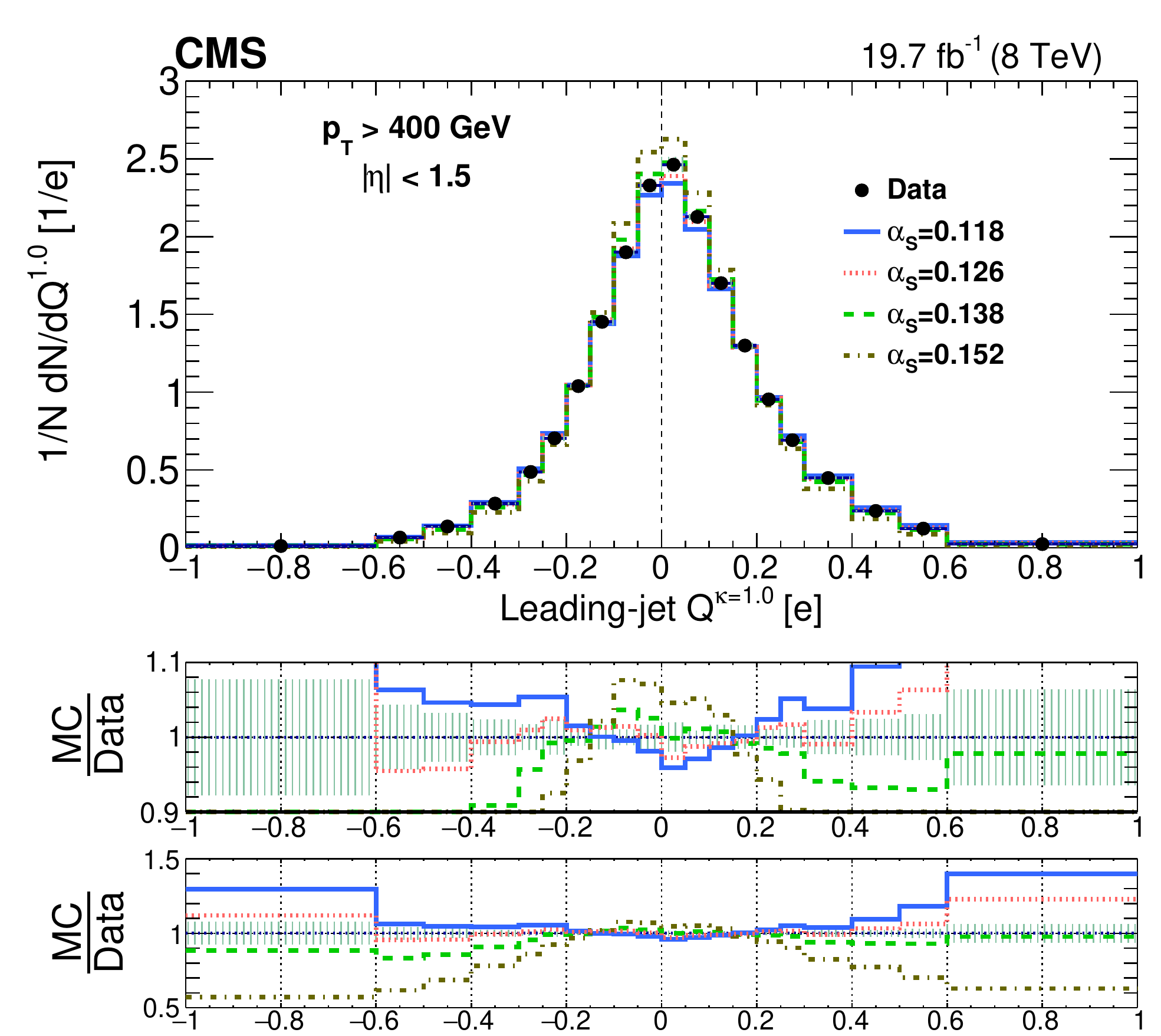}
\caption{\label{finalModelaS}Comparison of unfolded leading-jet charge distributions with predictions from \POWHEG+~\PYTHIA{8}.
The NLO \POWHEG prediction with the NLO CT10 PDF set is compared with predictions where the $\alpha_{S}$ parameter for final-state radiation in \PYTHIA{8} is varied from its default value of 0.138.
The default jet charge definition ($Q^\kappa$) for $\kappa=0.3$, 0.6, 1.0, the longitudinal jet charge definition ($Q_{L}^\kappa$), and the transverse jet charge definition ($Q_{T}^\kappa$) are shown.
Hashed uncertainty bands include both statistical and systematic contributions in data, added in quadrature.
The ratio of data to simulation is displayed twice below each plot with two different vertical scales.
}
\end{center}
\end{figure}

In Fig.~\ref{finalModelaS}, we vary the $\alpha_{S}$ parameter for the final-state radiation in \PYTHIA{8}, to which the jet charge distribution was found to be most sensitive, from its default value of 0.138.
This helps us to understand whether the underlying physics model in \PYTHIA{8} is in principle capable of simultaneously describing the effect observed in the various jet charge distributions.
All jet charge distributions, except $Q^{0.3}$, favor smaller values of $\alpha_{S}$ between 0.018 and 0.126 for the final-state radiation, while for $Q^{0.3}$ a larger value of $\alpha_{S}$ of around 0.158 is favored.
Therefore, we conclude that by varying the $\alpha_{S}$ parameter for the final-state radiation, the \POWHEG+~\PYTHIA{8} prediction can give an excellent description for most distributions, but not all of them with the same $\alpha_{S}$ parameter.
Thus specific jet charge distributions test aspects of the model that cannot be accommodated by a single parameter.

\section{Summary}

This paper presents measurements of jet charge distributions, unfolded for detector effects, with dijet events collected in proton-proton collisions at $\sqrt{s}=8\TeV$ corresponding to an integrated luminosity of 19.7\fbinv.
Distributions of the leading-jet charge are obtained for three ranges of leading-jet \pt and for three definitions of jet charge.
These three definitions of jet charge provide different sensitivities to parton fragmentation.
Three choices for the $\kappa$ parameter are considered, which provide different sensitivities to the softer and harder particles in the jet.
The variation of the jet charge with leading-jet \pt is sensitive to the quark and gluon jet content in the dijet sample.
In general, the predictions from \POWHEG+~\PYTHIA{8} and \POWHEG+~\HERWIG{++} generators show only mild discrepancies with the data distributions.
Nevertheless, the differences between the predictions from \POWHEG+~\PYTHIA{8} and \POWHEG+~\HERWIG{++} can be reduced with the help of these measurements.

\begin{acknowledgments}

We congratulate our colleagues in the CERN accelerator departments for the excellent performance of the LHC and thank the technical and administrative staffs at CERN and at other CMS institutes for their contributions to the success of the CMS effort. In addition, we gratefully acknowledge the computing centres and personnel of the Worldwide LHC Computing Grid for delivering so effectively the computing infrastructure essential to our analyses. Finally, we acknowledge the enduring support for the construction and operation of the LHC and the CMS detector provided by the following funding agencies: BMWFW and FWF (Austria); FNRS and FWO (Belgium); CNPq, CAPES, FAPERJ, and FAPESP (Brazil); MES (Bulgaria); CERN; CAS, MoST, and NSFC (China); COLCIENCIAS (Colombia); MSES and CSF (Croatia); RPF (Cyprus); SENESCYT (Ecuador); MoER, ERC IUT, and ERDF (Estonia); Academy of Finland, MEC, and HIP (Finland); CEA and CNRS/IN2P3 (France); BMBF, DFG, and HGF (Germany); GSRT (Greece); OTKA and NIH (Hungary); DAE and DST (India); IPM (Iran); SFI (Ireland); INFN (Italy); MSIP and NRF (Republic of Korea); LAS (Lithuania); MOE and UM (Malaysia); BUAP, CINVESTAV, CONACYT, LNS, SEP, and UASLP-FAI (Mexico); MBIE (New Zealand); PAEC (Pakistan); MSHE and NSC (Poland); FCT (Portugal); JINR (Dubna); MON, RosAtom, RAS, RFBR and RAEP (Russia); MESTD (Serbia); SEIDI, CPAN, PCTI and FEDER (Spain); Swiss Funding Agencies (Switzerland); MST (Taipei); ThEPCenter, IPST, STAR, and NSTDA (Thailand); TUBITAK and TAEK (Turkey); NASU and SFFR (Ukraine); STFC (United Kingdom); DOE and NSF (USA).

\hyphenation{Rachada-pisek} Individuals have received support from the Marie-Curie programme and the European Research Council and Horizon 2020 Grant, contract No. 675440 (European Union); the Leventis Foundation; the A. P. Sloan Foundation; the Alexander von Humboldt Foundation; the Belgian Federal Science Policy Office; the Fonds pour la Formation \`a la Recherche dans l'Industrie et dans l'Agriculture (FRIA-Belgium); the Agentschap voor Innovatie door Wetenschap en Technologie (IWT-Belgium); the Ministry of Education, Youth and Sports (MEYS) of the Czech Republic; the Council of Science and Industrial Research, India; the HOMING PLUS programme of the Foundation for Polish Science, cofinanced from European Union, Regional Development Fund, the Mobility Plus programme of the Ministry of Science and Higher Education, the National Science Center (Poland), contracts Harmonia 2014/14/M/ST2/00428, Opus 2014/13/B/ST2/02543, 2014/15/B/ST2/03998, and 2015/19/B/ST2/02861, Sonata-bis 2012/07/E/ST2/01406; the National Priorities Research Program by Qatar National Research Fund; the Programa Clar\'in-COFUND del Principado de Asturias; the Thalis and Aristeia programmes cofinanced by EU-ESF and the Greek NSRF; the Rachadapisek Sompot Fund for Postdoctoral Fellowship, Chulalongkorn University and the Chulalongkorn Academic into Its 2nd Century Project Advancement Project (Thailand); and the Welch Foundation, contract C-1845.
\end{acknowledgments}

\bibliography{auto_generated}

\cleardoublepage \appendix\section{The CMS Collaboration \label{app:collab}}\begin{sloppypar}\hyphenpenalty=5000\widowpenalty=500\clubpenalty=5000\textbf{Yerevan Physics Institute,  Yerevan,  Armenia}\\*[0pt]
A.M.~Sirunyan, A.~Tumasyan
\vskip\cmsinstskip
\textbf{Institut f\"{u}r Hochenergiephysik,  Wien,  Austria}\\*[0pt]
W.~Adam, E.~Asilar, T.~Bergauer, J.~Brandstetter, E.~Brondolin, M.~Dragicevic, J.~Er\"{o}, M.~Flechl, M.~Friedl, R.~Fr\"{u}hwirth\cmsAuthorMark{1}, V.M.~Ghete, C.~Hartl, N.~H\"{o}rmann, J.~Hrubec, M.~Jeitler\cmsAuthorMark{1}, A.~K\"{o}nig, I.~Kr\"{a}tschmer, D.~Liko, T.~Matsushita, I.~Mikulec, D.~Rabady, N.~Rad, B.~Rahbaran, H.~Rohringer, J.~Schieck\cmsAuthorMark{1}, J.~Strauss, W.~Waltenberger, C.-E.~Wulz\cmsAuthorMark{1}
% ...
\vskip\cmsinstskip
% ...
1:~~Also at Vienna University of Technology, Vienna, Austria\\
% ...

\end{sloppypar}
\end{document}